\documentclass[12pt]{article}
\usepackage[utf8]{inputenc}
\usepackage{amsfonts}
\usepackage{amsmath}
\usepackage{amssymb}
\usepackage{bigints}
\usepackage{booktabs}
\usepackage[nosort]{cite}
\usepackage{color}
\usepackage{dsfont}
\usepackage{float}
\usepackage{framed}
\usepackage{graphicx}
\usepackage{indentfirst}
\usepackage{mathrsfs}
\usepackage{multirow}
\usepackage{pdflscape}
\usepackage{setspace}
\usepackage{subdepth}
\usepackage{subfig}
\usepackage{titlesec}
\usepackage[dotinlabels]{titletoc}
\usepackage{wrapfig}
\usepackage[all]{xy}
\usepackage{young}
\usepackage[vcentermath]{youngtab}
\usepackage{relsize}
\usepackage{stackengine}
\usepackage{datetime}

\usepackage{datetime,color}
\usepackage{graphicx}
\usepackage{float}
\usepackage{authblk}

\newcommand{\be}{\begin{equation}}

\newcommand{\ee}{\end{equation}}

\numberwithin{equation}{section}
\DeclareFontShape{OT1}{cmr}{mx}{n}%
    {<->cmr10}{}
\newcommand{\mytitlefont}{\fontseries{mx}\selectfont}
\DeclareMathAlphabet{\titlemath}{OT1}{cmr}{mx}{n}

\begin{document}

\begin{titlepage}
\begin{flushright}
{\fontsize{14pt}{0pt} CPHT -RR086.092018 }
\end{flushright}

\begin{center}

~\\[2cm]

{\fontsize{20pt}{0pt} \mytitlefont Deforming black holes in AdS }

~\\[0.5cm]

{\fontsize{14pt}{0pt} Gary T. Horowitz $^{\sharp}$, Jorge E. Santos$^{\diamond}$, Chiara Toldo$^{\dag \star}$}

~\\[0.1cm]

\it{ ${}^{\sharp}$ Department of Physics,}\\ \it{University of California, Santa Barbara, CA 93106}

~\\[0.05cm]

\it{ ${}^{\diamond}$ Department of Applied Mathematics and Theoretical Physics,}
\\ \it{University of Cambridge, Wilberforce Road, Cambridge CB3 0WA, UK}

~\\[0.05cm]

\it{ ${}^{\dag}$ Kavli Institute for Theoretical Physics}\\ \it{University of California, Santa Barbara, CA 93106}

~\\[0.05cm]

\it{ ${}^{\star}$ Centre de Physique Th\'eorique  (CPHT), Ecole Polytechnique, \\
	91128 Palaiseau Cedex, France }

~\\[0.8cm]

\end{center}

  \vspace{60pt}

\noindent 
We  investigate how changes in the boundary metric affect the shape of AdS black holes. Most of our work is analytic and based on the AdS C-metric. Both asymptotically hyperbolic and compact black holes are studied. It has recently been shown that the AdS C-metric contains configurations of highly deformed  black holes, and we show that these deformations are usually the result of  similar deformations of the boundary metric. However, quite surprisingly, we also find cases where the horizon is insensitive to certain large changes in the boundary geometry. This motivates the search for a new family of black hole solutions with the same boundary geometry in which the horizon does respond to the changes in the boundary. We numerically construct these solutions and  we (numerically) explore how the horizon response to boundary deformations depends on temperature.

\vfill 
    
    \noindent
  
  \end{titlepage}
  
   \newpage
  
  \tableofcontents
    
  \section{Introduction}
  
Motivated by gauge/gravity duality, there has been considerable interest in understanding the  asymptotically anti-de Sitter (AdS) versions of gravitating objects such as black holes.  The study of black holes in AdS is much richer than the  familiar asymptotically flat case. In particular, four dimensional static vacuum black holes in AdS can have horizons of various shape: known solutions include noncompact planar and hyperbolic horizons, as well compact horizons of arbitrary genus $g$. The curvature of the horizon need not be constant, and can include various deformations \cite{Chen:2015zoa}. There are even  noncompact horizons with finite area, where a spherical region is connected to the boundary by an infinitely long spike \cite{Chen:2016rjt}\footnote{Adding matter fields, one can find black holes with two spikes like those considered in \cite{Gnecchi:2013mja,Klemm:2014rda}. In the body of this paper we will consider only vacuum solutions. In an Appendix, we briefly discuss the effects of adding a Maxwell field.}.

The reason for the wide variety of black holes in AdS is that the boundary at infinity is timelike and one is free to specify a boundary metric. Deforming the boundary metric of a spacetime containing a black hole typically deforms the black hole horizon.  This has been proven mathematically  \cite{Anderson:2002xb} and studied numerically (see, e.g.,\cite{Markeviciute:2017jcp,Blazquez-Salcedo:2017kig}). The main aim of this paper is to investigate the effects of such boundary deformations on the horizon using an exact analytic solution. This will provide  new insight into the response properties of  black hole horizons to boundary changes. We will encounter a couple of puzzles which we will  resolve through numerical work.

The exact solution we will use is the AdS C-metric. The original C-metric appeared in the studies of Levi-Civita \cite{levicivita1917} and Weyl \cite{weyl1917} around 1917. The generalization to include a cosmological constant was given in \cite{Plebanski:1976gy} and (for a certain range of parameters) interpreted as   describing a pair of accelerated black holes in AdS  \cite{Podolsky:2002nk,Dias:2002mi,Krtous:2005ej}. This solution has found many other applications including  describing black holes living on branes in brane-world scenarios \cite{Emparan:1999wa,Emparan:1999fd} and also describing  strongly coupled Hawking radiation  \cite{Hubeny:2009ru,Hubeny:2009kz}. 

\begin{figure}[H]
   \begin{center}
{\includegraphics[width=0.3\textwidth]{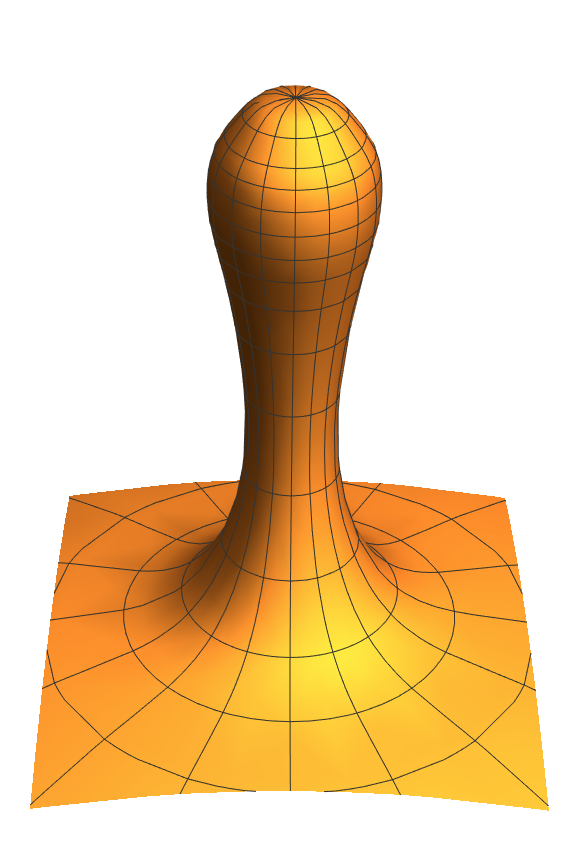} \caption{Embedding in three-dimensional hyperbolic space of a ``black globule", i.e., a protrusion on the hyperbolic black hole horizon resembling a spherical mass suspended over the black hole horizon. \label{Fig0}}}
\end{center}
\end{figure}

It was recently shown  that in a region of parameter space, a sub-class of the AdS C-metric described black holes whose horizon asymptotically approaches (a quotient of) the two-dimensional hyperbolic plane and contains a deformation at its center \cite{Chen:2015zoa}. For a certain choice of parameters the deformation of the horizon can be quite drastic: in the central region the curvature can become positive and the protrusion assumes a spherical form emerging from the asymptotically hyperbolic horizon (see Fig.~\ref{Fig0}). The neck connecting the spherical part to the rest of the horizon can be made arbitrarily thin and due to the resemblance to a drop of liquid these configurations were dubbed ``black globules".

In this paper we investigate the boundary geometry of these spacetimes to see the cause of such deformations. As expected, in many cases we find that the boundary geometry is deformed in a way that mimics the horizon geometry, providing a simple explanation for the behavior of the horizon.  However, there is a  regime where the boundary geometry grows by an arbitrarily large amount and the horizon geometry is unaffected! 
Since the solutions of \cite{Chen:2015zoa} are exact and in a closed form, we can show this analytically. This is not only surprising but is in tension with gauge/gravity duality: A thermal state of a QFT living on the boundary geometry will have an entropy that is proportional to the area of the boundary. This will not agree with the area of the black hole horizon\footnote{This is true after regulating the infinite area associated with the asymptotic region.}. This strongly suggests that these solutions do not dominate the canonical ensemble, and there is another family of black hole solutions with the same boundary geometry which is dual to a thermal state.

We then turn to numerical work and indeed find a new class of black hole solutions with the same boundary metric. These new horizons do increase their area in response to the large increase in the boundary area. We also study the temperature dependence of the horizon response to boundary deformations, both for the C-metric and the new solutions. This is done numerically in both cases since the C-metric provides only a single temperature solution for each boundary geometry. Curiously, the horizon response  appears to be greatest near the temperature of the C-metric. 

In addition, we compute the boundary stress tensor and total energy of these solutions. Even though the black hole and boundary are both asymptotically hyperbolic, the C-metric solutions have finite total energy. This is because they have temperature $T = 1/(2\pi)$, and the uniform hyperbolic black hole with this temperature is locally AdS with zero energy.

It turns out that for a certain range of parameters,  the C-metric has static timelike geodesics. This suggests that one can add hovering black holes: small spherical black holes that hover above the asymptotically hyperbolic horizon.

Finally, we extend our study of the C-metric to include boundary metrics that contain a black hole (so-called ``black droplet" solutions), and boundary metrics that are compact. In the latter case we again encounter a range of parameters in which the area of the compact boundary grows by an arbitrarily large amount while the area of the spherical black hole remains unaffected. 
As before, this strongly suggests that there is another family of solutions with the same boundary geometry.  We then again turn to numerical work and find these new solutions. Unlike the C-metric black holes which have negative specific heat, the new solutions have positive specific heat and grow in size when the boundary grows.

The boundary metric has a conical singularity and both the solutions described by the C-metric and the new solutions we find  contain a conical singularity in the bulk along one axis representing a string pulling on the black hole. As we will discuss, it is likely that the solution that dominates the canonical ensemble in this case will be a smooth bulk solution with the same boundary metric. Finding such a solution remains an open problem.  
  
 After a review of known results concerning the AdS C-metric and its domain structure, we describe in some detail the deformations of the horizon (section 3), noticing that the globule configurations of \cite{Chen:2015zoa} appear in a more extended phase space than previously noticed. In section 4 we study the boundary geometry that causes these deformations and   show the (non-)response of the horizon to some large boundary deformations. We then describe our new solutions which are found numerically, and some of their properties. Next, we discuss the case where the boundary metric contains a black hole. Finally we consider the case of compact boundaries. We again show the (non-)response of the horizon to large boundary deformations, and numerically construct a new class of black holes which do respond.  In an Appendix we discuss the charged generalization of the C-metric. Once again, there are regions of parameters where the boundary undergoes large changes and the black hole horizon does not respond.

  \section{Setup: the AdS C-metric}
  \subsection{Analytic solution}
  
 A new form for the AdS C-metric with a cosmological constant was proposed in \cite{Chen:2015vma}, which generalizes the one originally found in \cite{Plebanski:1976gy}:
   \be \label{C-metric}
ds^2 = \frac{l^2}{(x-y)^2} \left( F(y) \mathrm{d}t^2 -\frac{\mathrm{d}y^2}{F(y)} +\frac{\mathrm{d}x^2}{G(x)} +G(x) \mathrm{d}\phi^2 \right)
\ee
where the structure functions $F(y)$ and $G(x)$ are
\be \label{funch_h}
F(y) = y[1+ \nu +(\mu+\nu) y + \mu y^2] \,,
\ee
\be \label{funcg_h}
G(x)= (1+x) ( 1 + \nu x + \mu x^2)\,,
\ee
and $\mu $ and $\nu$ are real parameters. (We follow the conventions of \cite{Chen:2015vma,Chen:2015zoa}.) This static, axisymmetric metric satisfies the Einstein's equation with a cosmological constant 
\be
R_{ab} =\Lambda g_{ab}
\ee
with $ \Lambda= -3/l^2$. 
Notice the relation
\be
G(x) - F(x) =1\,,
\ee
which will become useful later.

The roots of the structure function $G(x)$ are
\be
x_0 =-1, \qquad x_{\pm} = \frac{-\nu \pm \sqrt{\nu^2-4\mu}}{2 \mu}
\ee
and those of $F(y)$ are
\be
y_0=0\,, \qquad y_{\pm} = \frac{-(\mu+\nu)\pm \sqrt{(\mu-\nu)^2 -4 \mu}}{2 \mu}\,.
\ee
Zeros of $F(y)$ correspond to horizons, while those of $G(x)$ correspond to axes of the rotation symmetry. To avoid a conical singularity along the $x=-1$ axis, we periodically identify $\phi$ with period 
\be
\Delta \phi =  \frac{4\pi}{|G'(-1)|} = \frac{4\pi}{|1+\mu-\nu|}
\ee

We will be interested in regions of spacetime containing a horizon, an axis of symmetry, and an asymptotic infinity, where the latter lies at $x=y$. For this reason we will restrict the range of the coordinates $x,y$ to lie between the lines $x=x_0$, $y=y_0$ and the line $x=y$. The maximum coordinate range is therefore
\be \label{range}
-1 \le x <y\le0\,.
\ee
The Kretchmann scalar of the solution is
\be
R^{\mu \nu \rho \sigma} R_{\mu \nu \rho \sigma} = \frac{24}{l^4} + \frac{12 \mu^2 (x-y)^6}{l^4}
\ee
One can see that it is bounded in the range \eqref{range}, hence there are no curvature singularities in the region of spacetime that we will consider. 

When $\mu = 0$, the Kretchmann scalar is constant and the solution reduces to pure AdS. This is easy to see when $\nu = -1$ since then 
\be
F(y) = -y^2\,,
\qquad
G(x)= (1+x) ( 1 - x)\,,
\ee
and \eqref{C-metric}  can be recast in the form
\be
\mathrm{d}s^2 = \frac{l^2 \mathrm{d}r^2}{r^2} + r^2 \left( -\mathrm{d}t^2 + \mathrm{d}\rho^2 + \rho^2 \mathrm{d}\phi^2 \right)
\ee
 by means of the coordinate transformation 
\be
r = \frac{l y}{(x-y)},  \qquad \rho =  \frac{\sqrt{1-x^2}}{y}
\ee
The choice $\mu =0, \nu = -1$ corresponds to $\lambda =0$ and $\kappa =-1$ in \cite{Emparan:1999fd}.

To have the  correct signature $(-+++)$, with timelike Killing vector $\partial_t$, we require  
\be
F(y) < 0 \qquad \text{and} \qquad G(x)>0
\ee
in the coordinate range \eqref{range}. This gives restrictions on the parameters $\mu$ and $\nu$ which is summarized in the next subsection. In the rest of the paper we set $l=1$ for simplicity.

  \subsection{Domains}
  
  The parameter space of these solutions was studied in \cite{Chen:2015zoa} and we summarize here  the main results. There are four regions where the solutions have the correct Lorentzian signature in the range \eqref{range}. 
  
\begin{figure}[H]
{\includegraphics[width=0.9\textwidth]{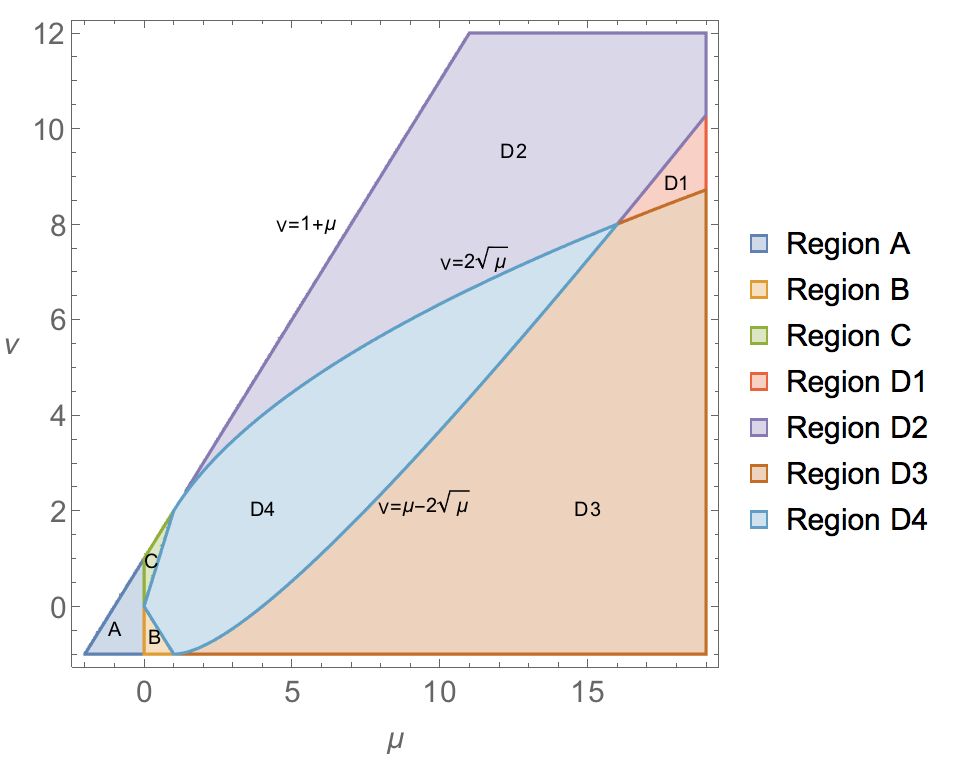} \caption{\label{Fig1} Plot of the ($\mu, \nu$) parameter space for regions A-D. } }
\end{figure}

  \begin{itemize}
  \item \textbf{Region A}: This  region is characterized by $\mu<0$. Lorentzian signature requires that $x_0$ and $y_0$ are middle roots, which leads to the ordering 
  \be
  y_+<x_+<-1<0<y_-<x_- \,,
  \ee
  The $\mu,\nu$ parameter range is shown in blue in Figure \ref{Fig1}, and it corresponds to the region
  \be
  \nu< 1+ \mu \quad \text{ and} \quad \nu > -1 \,.
  \ee

  \item \textbf{Region B:} This region is characterized by $\mu>0$ and the fact that $x_0$ and $y_0$  are both the smallest roots. The ordering of the roots is thus
  \be
  -1<0<y_-<x_-<x_+< y_+
  \ee
  and the region is depicted in orange in Fig. \ref{Fig1}, corresponding to
  \be
  \nu>-1 \quad \text{and} \quad \nu < -\mu \,.
  \ee
  \item \textbf{Region C:} This region is characterized by $\mu>0$ and by the fact that $x_0 $ and $y_0$ are both the largest roots. It is depicted in green in the Fig. \ref{Fig1} with
  \be
  \nu<1+\mu \quad \text{and} \quad \nu>2\mu\,,
  \ee
  and the ordering is
  \be
x_-<y_-<y_+<x_+<-1<0
  \ee
  \item \textbf{Region D:}  This region is characterized by $\mu>0$ and by the fact that  $x_0$ and $y_0$  bound all the other roots. The ordering is
  \be \label{ord}
  -1<y_-<y_+<x_-<x_+<0\,.
  \ee
  This region of $\mu,\nu$ parameter space is the largest of the four, it is bounded by
  $\nu < 2 \mu$, $ \nu>-\mu$, $ \nu>-1$ and $\nu<1+\mu$, and is  divided into four subregions D1-D2-D3-D4  in Fig. \ref{Fig1} which we now explain.  
    \end{itemize}

\begin{figure}[H]
\begin{centering}
{\includegraphics[width=0.88\textwidth]{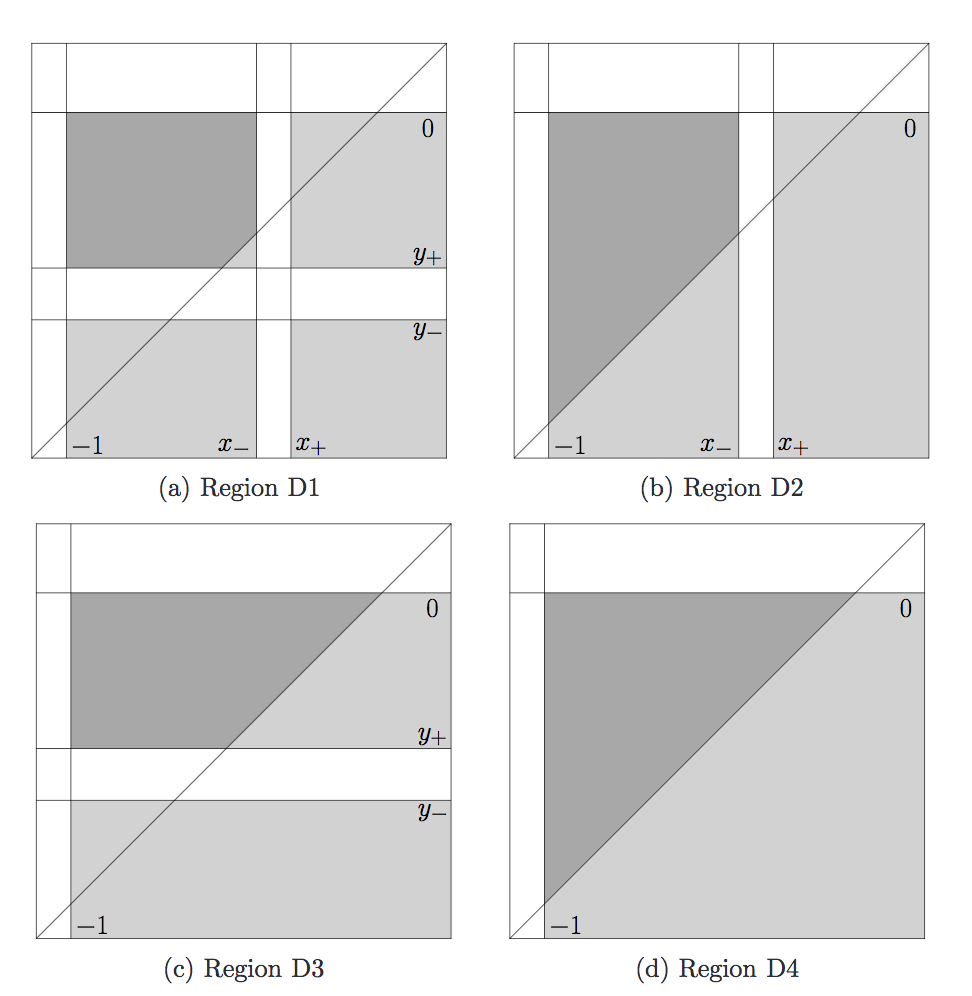}  \caption{The range of $x,y$ coordinates. Shaded areas correspond to the coordinate ranges with Lorentz signature. Darker shaded regions are of most interest. Figure { reproduced with permission from Teo et al. \cite{Chen:2015zoa}.}
\label{Fig9}}}
\end{centering}
\end{figure}

In region A,  all the roots are real. However, in the other regions, it is possible that $x_\pm$ and $y_\pm$ become complex. Recall that we always have one axis at $x=-1$ and one horizon at $y=0$.  If $x_\pm$ are real, there is another axis, and our black holes are spherical. Since we have already fixed the periodicity of $\phi$ so the $x=-1$ axis is smooth, there will typically be a conical singularity along the second axis. This can be interpreted as a cosmic string pulling on the black hole. If $y_\pm$ are real, there is another horizon in the spacetime which we will interpret below.  
\noindent Focussing on region D, there are two lines, 
\be
\nu = 2 \sqrt{\mu} \qquad \text{and} \qquad \nu=\mu-2\sqrt{\mu}\,,
\ee
which separate the various regions. In particular, with reference to Fig. \ref{Fig9}:
\begin{itemize}
\item Region D1 is characterized by both $x_{\pm}$ and $y_{\pm}$ real, and its region is bounded below by $ 2 \sqrt{\mu} < \nu$ and above by $\nu < \mu - 2\sqrt{\mu}$.   Its $x-y$ diagram is a pentagon. These solutions describe spherical black holes that are accelerating, and the second horizon  at $y=y_+$ is an acceleration horizon. 

\item Region D2 has two real $x_{\pm} $ roots, while $y_{\pm}$ are complex. It is characterized by $\nu>2 \sqrt{\mu}$ and $\nu>\mu-2\sqrt{\mu}$. This region has a trapezoid domain and describes an accelerating spherical black hole without an acceleration horizon. One can interpret it as a spherical black hole in AdS which is held at a fixed distance away from the center. 
\end{itemize}

\begin{itemize}
\item Region D3 has two real $y_{\pm} $ roots, while $x_{\pm} $ are complex. It is characterized by $\nu<2 \sqrt{\mu}$ and $\nu < \mu-2\sqrt{\mu}$. This region is again characterized by a trapezoidal domain. However, there are major differences with respect to region D2. Indeed in this case the horizon at $y=0$ is noncompact and describes a (deformed) hyperbolic black hole.  The second horizon at  $y = y_+$ is a horizon for a black hole which reaches  the boundary itself. This describes a   \textit{black droplet } \cite{Hubeny:2009ru,Hubeny:2009kz}, where a black hole anchored on the boundary is ``suspended" over a deformed hyperbolic black hole.

\item Region D4 has complex $x_{\pm}$ and $y_{\pm}$. It is characterized by $\mu-2\sqrt{\mu} < \nu < 2\sqrt{\mu}$.  The only real roots of $F(y)$ and $G(x)$ are $y=0$ and $x=-1$. This region has a triangular domain and it represents a deformed hyperbolic black hole. 

\end{itemize}
  
  \section{Deforming the horizon}
  
 We start in region D4 shown in Fig. \ref{Fig9}d which has one horizon at $y=0$, and one axis of symmetry at $x=-1$.
  The diagonal edge $x=y$ represents asymptotic infinity and runs from $-1$ to $0$. The Hawking  temperature (associated with $\partial_t$) of the
 horizon is 
\be\label{hawkT0}
T =\frac{F'(0)}{4\pi} = \frac{1+\nu}{4\pi}
\ee

Let us now study the horizon geometry. The induced metric on the horizon, which is located at $y=0$, takes the form
\be \label{metric_horizon}
\mathrm{d}s^2 = \frac{1}{x^2} \left( \frac{\mathrm{d}x^2}{G(x)} + G(x) \mathrm{d}\phi^2 \right)
\ee
The scalar curvature ${\cal R}$ of (\ref{metric_horizon})  is
\be\label{horcurv}
{\cal R} = -2 (1+ \mu x^3)\, 
\ee
Notice that the value of the curvature  depends only on $\mu$. But this does not mean that the horizon geometry is independent of $\nu$. The proper distance from the axis $x=-1$ to any point $x>-1$ depends on $\nu$.

Since the horizon is at $y=0$, $x\to 0$ represents the asymptotic region. In the limit $x \rightarrow 0$ the warp factor $G(x)$ approaches 1, hence the horizon \eqref{metric_horizon} approaches (a quotient of) two-dimensional hyperbolic space. One can also see this from ${\cal R}$, which approaches the curvature of a unit hyperboloid as 
 $x \rightarrow 0$. 

However, something unusual can happen near the axis $x = -1$ \cite{Chen:2015zoa}. The horizon can have positive curvature there provided 
\be
\mu>1\,.
\ee
If the region of positive curvature is large enough,
  the horizon geometry becomes approximately spherical in the region around $x=-1$ and a round  protrusion  out of the asymptotically  hyperbolic horizon appears. 
  
 These configurations of deformed hyperbolic black holes were discovered in \cite{Chen:2015zoa} for region D4 and dubbed ``black globules" for their resemblance to drops of liquid. We will denote with this term configurations of black holes whose $g_{\phi\phi}$ is not monotonic, so that indeed the protrusion resembles a spherical mass being pulled out of  the black hole horizon.
 
\begin{figure}[H]
{\includegraphics[width=0.99\textwidth]{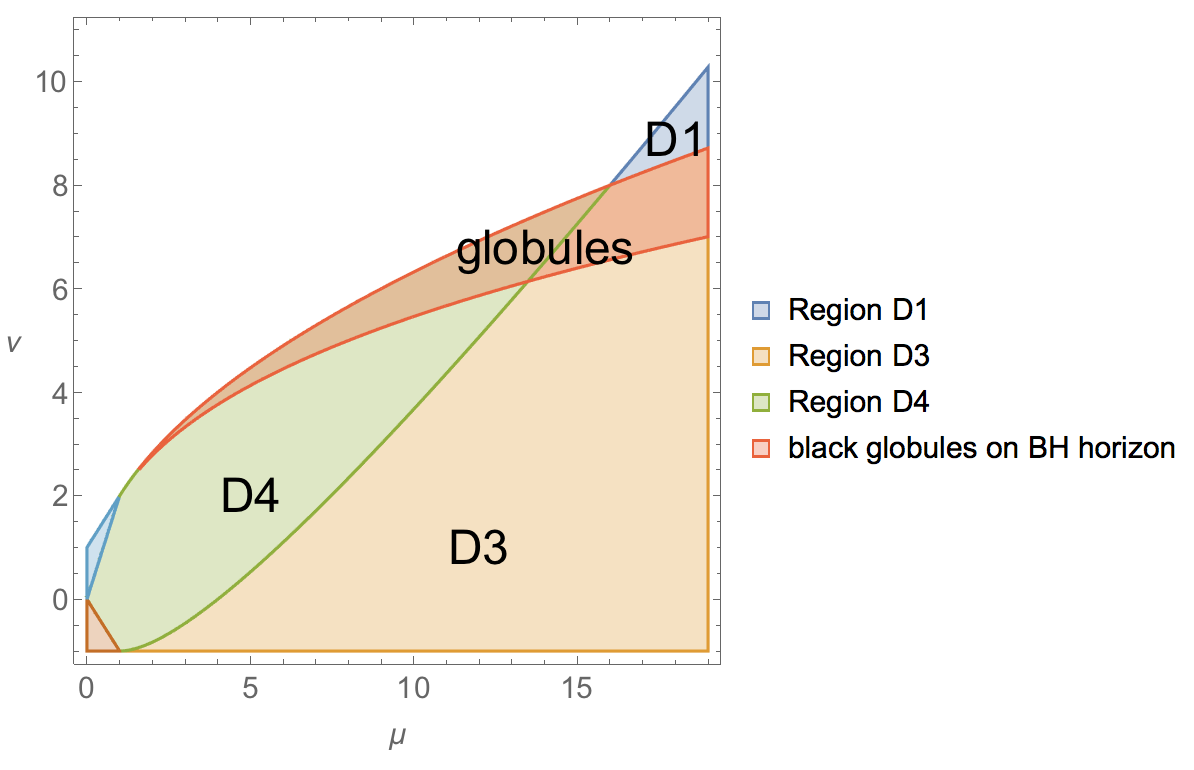} \caption{Plot of the ($\mu,\nu$) parameter space
with the region of black globules denoted  in red. Note that it includes part of regions D3 and D4. \label{Fig3}}}
\end{figure}

We can map the parameter region where black globules are present. It turns out that $\sqrt{G(x)}/x$ is not monotonic (and is defined throughout the entire interval $-1<x<0$) if 
\be
-1+ 3 \mu^{1/3} < \nu < 2 \sqrt{\mu}\,,
\ee
which marks the region of  black globules. This region is denoted in red in Fig \ref{Fig3}. Notice that it includes solutions in both regions D3 and D4.\footnote{It was  claimed in \cite{Chen:2015zoa} that black globules can only exist in region D4.}

\vspace{3mm}

\begin{figure}[H]\begin{center}
{\includegraphics[width=0.85\textwidth]{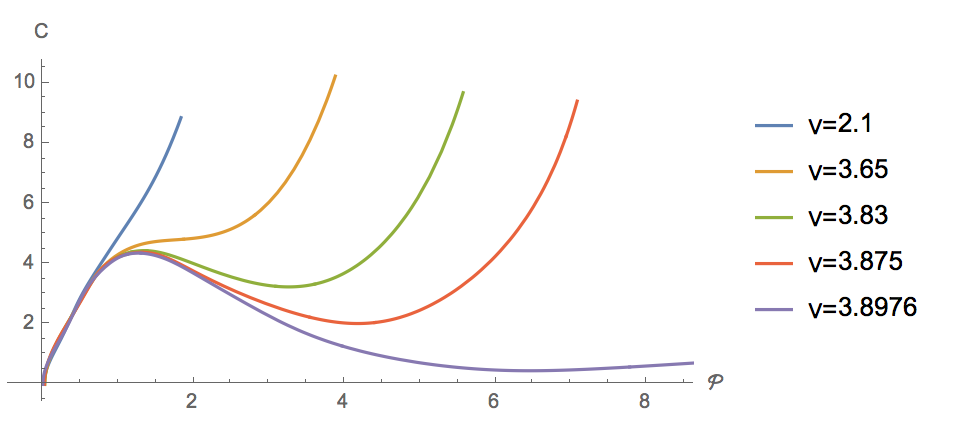}  \caption{Deformed hyperbolic black holes can become black globules. The circumference of the $\phi $ circles on the horizon is shown as a function of proper distance $\mathcal{P}$ from the axis for $\mu = 3.8$ and various $\nu$, approaching the border with D2 at $\nu \sim 3.899$. In this specific example, the bottom three curves (green, red and purple lines) correspond to black globules.
\label{Fig2}}}
\end{center}
\end{figure}

By tuning the parameters $\mu, \nu$ one can make the protrusion more prominent  or the ``neck" of the geometry more narrow. 
Notice that, if we approach the limiting value $\nu = 2 \sqrt{\mu}$ dividing region D4 from region D2, the neck becomes infinitely long. This is because $G(x)$ develops a double zero  when $\nu = 2 \sqrt{\mu}$ making $x = x_+ (= x_-)$ infinitely far away from the axis $x= -1$.  Since the length of the $\phi$ circles goes to zero as $x \to x_+$, the horizon geometry resembles an infinite spike. Increasing $\nu$ further, we enter D2 where the black hole {horizon has spherical topology}. This transition is illustrated in Fig \ref{Fig2} where we plot the circumference of the $\phi$ circles as a function of proper distance from the axis.

\begin{figure}[H] \begin{center}
{\includegraphics[width=0.85\textwidth]{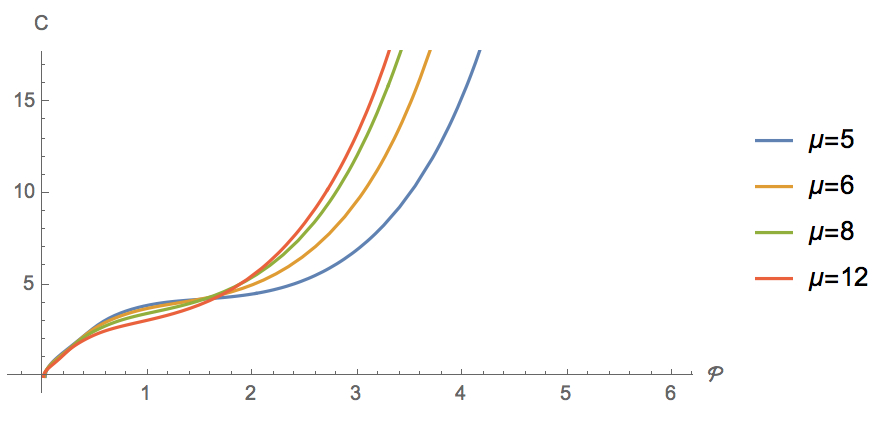} \caption{Deformed hyperbolic black holes for fixed $\nu=4$. The values of $\mu$ shown include points in both D3 and D4. \label{Fig4}}}
\end{center}
\end{figure}

The transition between D4 and D3 is much smoother for the horizon. 
In Fig. \ref{Fig4} the circumference of the $\phi $  circles on the horizon  for different values of $\mu$ is shown. This includes solutions in both D4 and D3.

The key question is what is causing these deformations of  the horizon. The only possible source is the metric on the conformal boundary at infinity. So we now turn to 
examine the boundary geometry.

\vspace{4mm}

\section{Boundary metrics}
\subsection{Region D4 \label{RegD4}}

\vspace{5mm}

We must first decide which conformal frame to use to describe the boundary geometry. Since the solutions are static and axisymmetric, it is natural to choose a frame for which the boundary geometry is ultrastatic, \emph{i.e.} it takes the form:
\be
\mathrm{d}s^2_{\partial} =-\mathrm{d}\tilde{t}^2 + \mathrm{d}r^2 + f(r) \mathrm{d}\phi^2\,.
\label{eq:boundarymetric}
\ee
Note that we are anticipating the time coordinate $t$ appearing in the original line element \eqref{C-metric} will not be the one used asymptotically (which we denote by $\tilde{t}$).

To obtain a boundary metric of this form, we start by pulling out a factor of $-F(y)$, and writing the metric \eqref{C-metric} as
   \be \label{C-metric2}
\mathrm{d}s^2 = -\frac{ F(y)}{(x-y)^2} \left( - \mathrm{d}t^2 +\frac{\mathrm{d}y^2}{F(y)^2} - \frac{\mathrm{d}x^2}{F(y)G(x)} -\frac{G(x)}{F(y)} \mathrm{d}\phi^2 \right)
\ee
Since the boundary corresponds to $x=y$, the induced metric can be obtained by setting $x=y$ in the expression in parenthesis:
\be
\widetilde{{\mathrm{d}s}}_\partial^2 = -\mathrm{d}t^2 +\frac{\mathrm{d}x^2}{F^2(x)G(x)} - \frac{G(x)}{F(x)} \mathrm{d}\phi^2 
\label{eq:bdy0}
\ee
Although this metric has the desired ultrastatic form and is asymptotically hyperbolic, it does not approach a hyperboloid of unit radius. This can be seen by taking the limit $x \rightarrow 0$ using the leading behavior $G =1, F = (1+\nu)x$ and setting $x = -(1+\nu)e^{-2r}$ to obtain
\be
\widetilde{\mathrm{d}s}^2_\partial = -\mathrm{d}t^2 + \frac{4}{(1+\nu)^2}\left[\mathrm{d}r^2 + \frac{e^{2r}}{4} \mathrm{d}\phi^2\right]
\ee
The quantity in brackets is a unit radius hyperbolic plane (since $\sinh^2 r \to e^{2r}/4 $ for large $r$). Since we want to keep the asymptotic geometry on the boundary fixed as we vary $\nu$, we will do an additional constant conformal transformation and set $\tilde t = (1+\nu) t/2 $ to obtain
\be
\mathrm{d}s_\partial^2 = -\mathrm{d}\tilde t^2 +\frac{(1+\nu)^2}{4} \left[\frac{\mathrm{d}x^2}{F^2(x)G(x)} - \frac{G(x)}{F(x)} \mathrm{d}\phi^2 \right] \equiv -\mathrm{d}\tilde t + \mathrm{d}r^2 + f(r) \mathrm{d} \phi^2
\label{withfactor}
\ee
The rescaling of the time coordinate induces a change in normalisation of the static bulk Killing field from $\partial/\partial {t}$ to $\partial/\partial {\tilde{t}}$, which means the bulk temperature is now simply given by
\be
T = \frac{1}{2\pi}\,.
\label{eq:tempD4}
\ee

\begin{figure}[H] \begin{center}
{\includegraphics[width=0.7\textwidth]{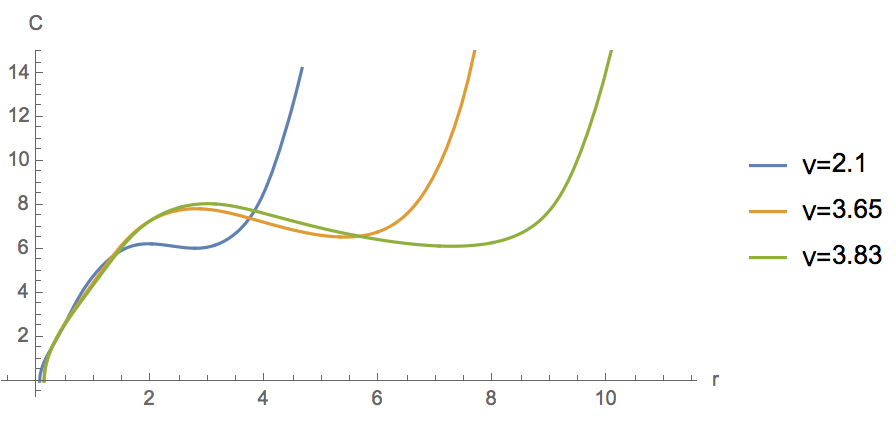} 
\caption{The boundary geometry as a function of the proper distance with $\mu = 3.8 $ and various values of $\nu$. \label{D4bdy}}}
\end{center}
\end{figure}

We can now study how the spatial geometry depends on $\mu$ and $\nu$. Figure \ref{D4bdy} shows the circumference of the $\phi$ circles as a function of $r$ for a constant  $\mu = 3.8$, and varying $\nu$. The fact that $C$ is not monotonic shows that the boundary geometry can resemble a  globule also. In fact,
 whenever there is a globule on the horizon in region D4, there is a corresponding globule in the boundary itself (see  Fig. \ref{Fig2}). It is this deformation of the boundary that induces a response in the black hole horizon. 
  
\begin{figure}[H] \begin{center}
{\includegraphics[width=0.31\textwidth]{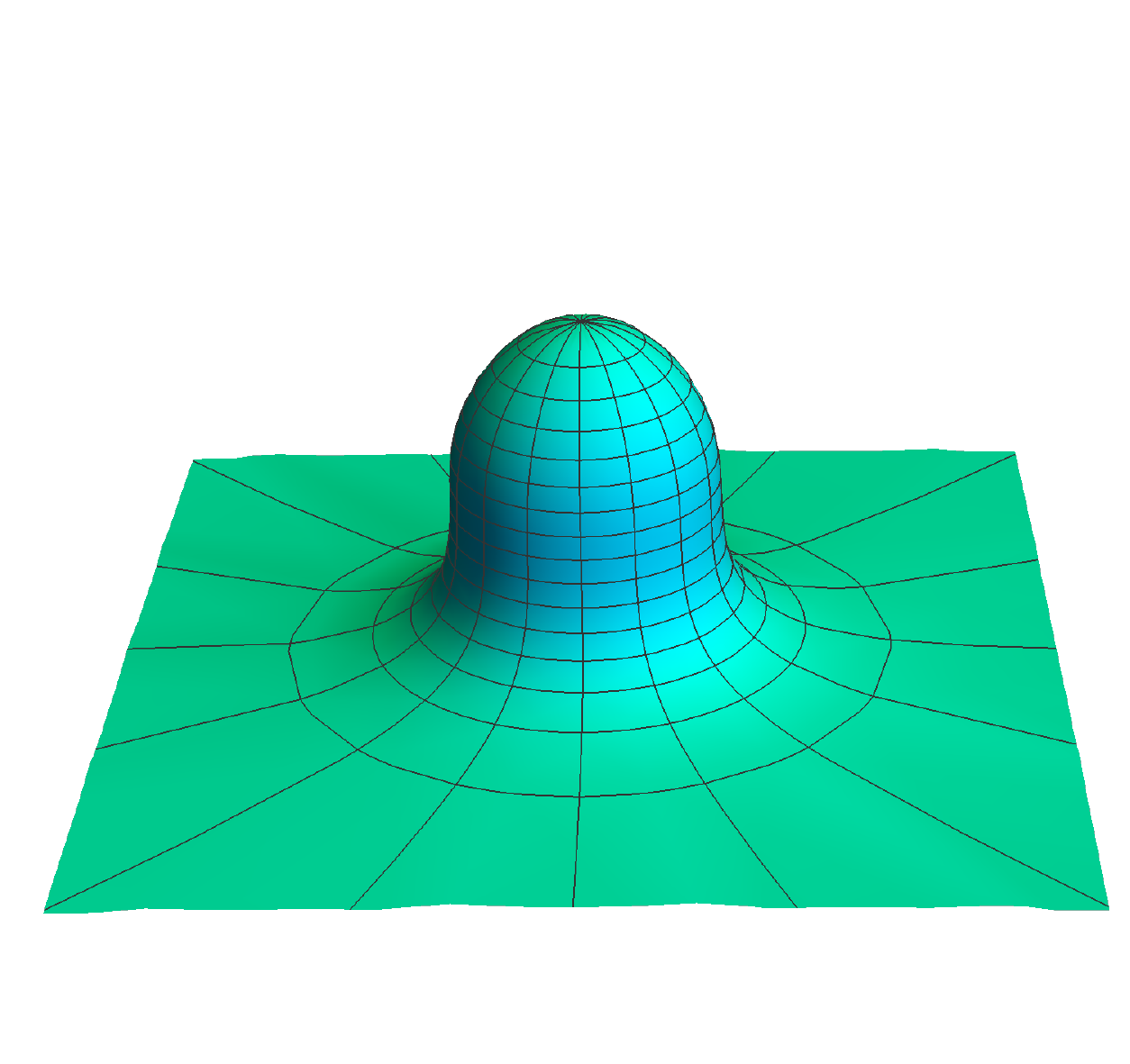}\includegraphics[width=0.369\textwidth]{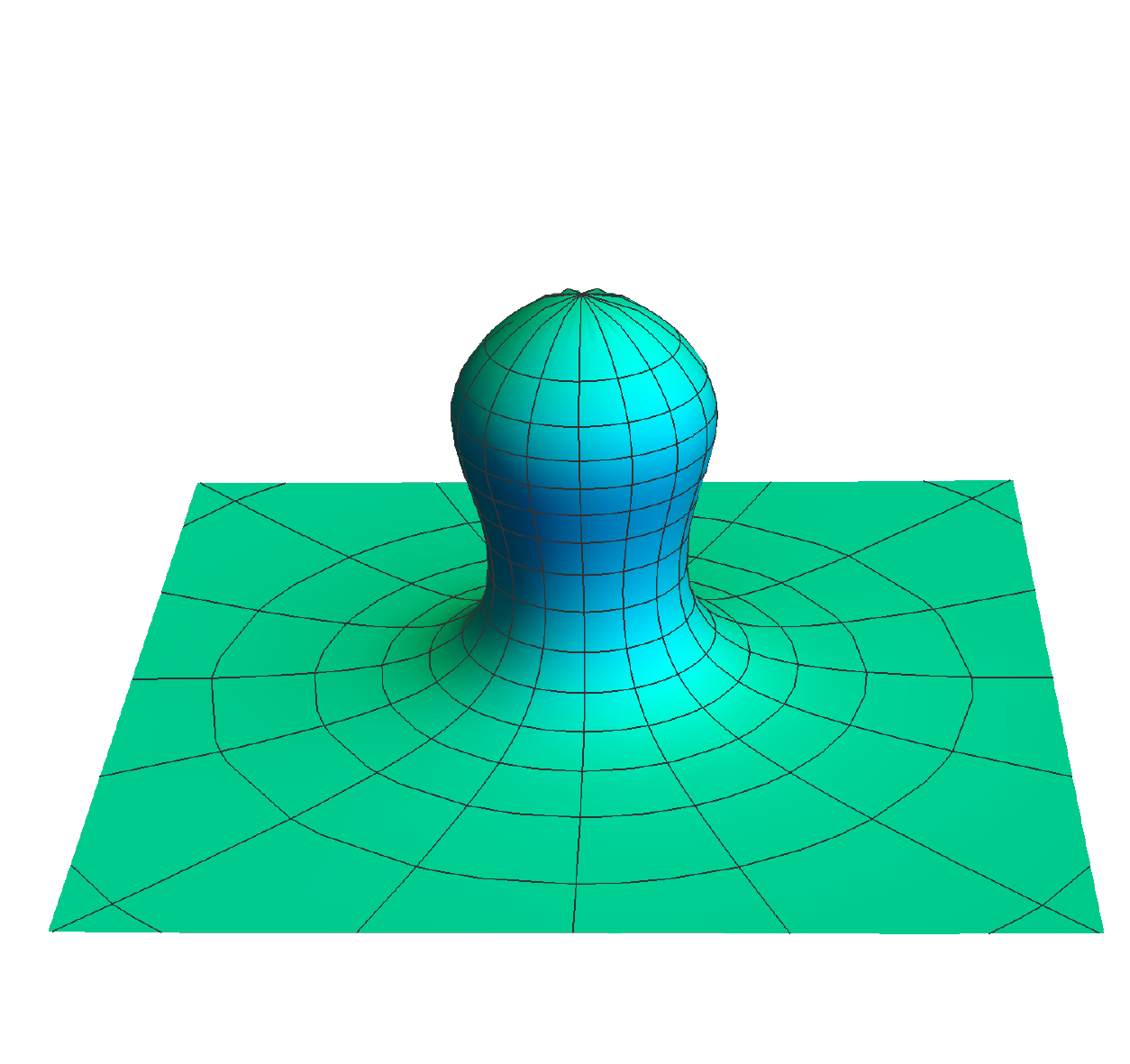} \includegraphics[width=0.312\textwidth]{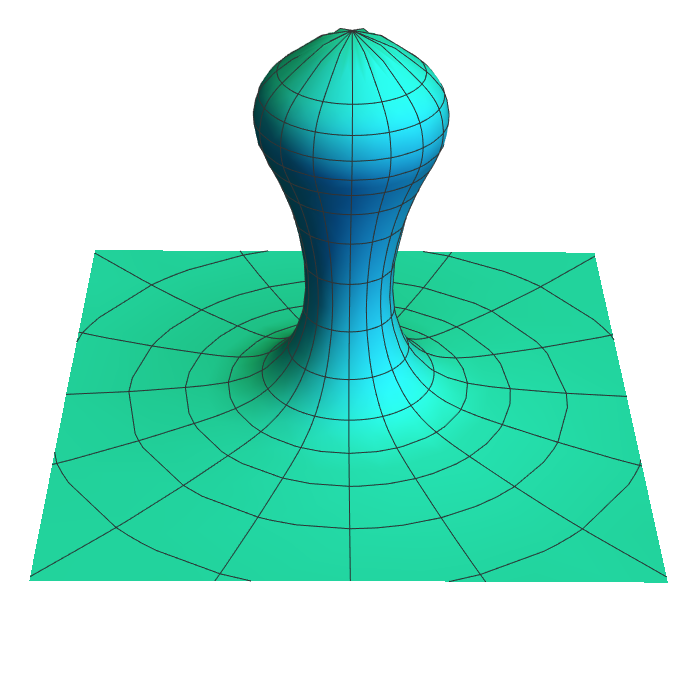} \\
\includegraphics[width=0.32\textwidth]{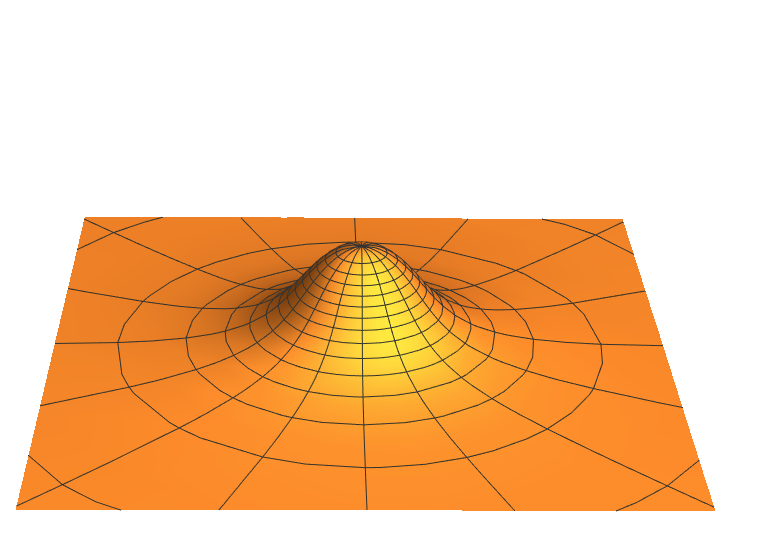}
\includegraphics[width=0.275\textwidth]{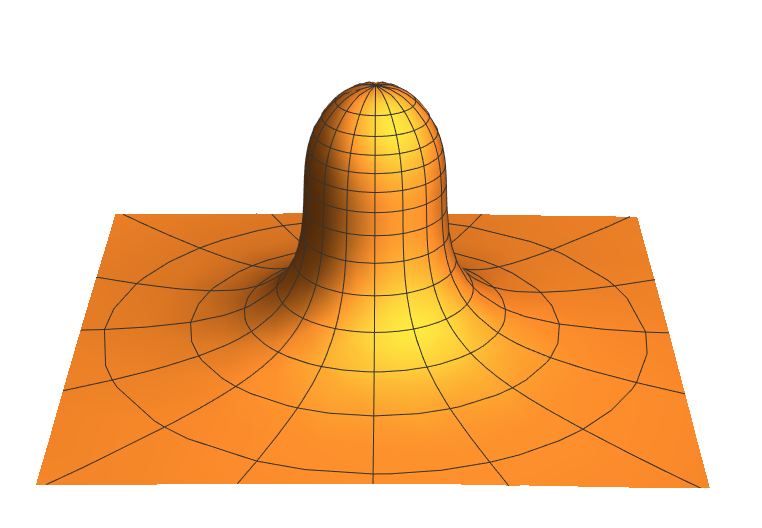}\includegraphics[width=0.245\textwidth]{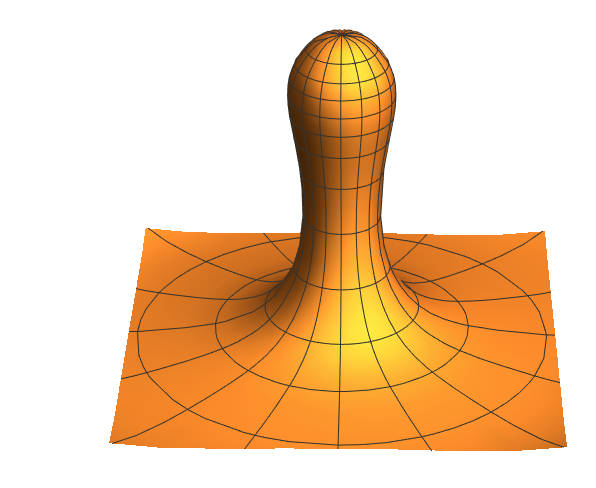} 
\caption{The embeddings of the boundary geometry in  hyperbolic space are shown in light blue in the upper panel. The corresponding horizon geometry is shown in yellow. In the plots, $\mu = 3.8$ and the values of $\nu$, from left to right are  $\nu= 2.1 $, $\nu= 3.65$, $\nu= 3.83$. \label{Fig5}}}
\end{center}
\end{figure}

\begin{figure}[H]
  \begin{center}
{ \includegraphics[width=0.52\textwidth]{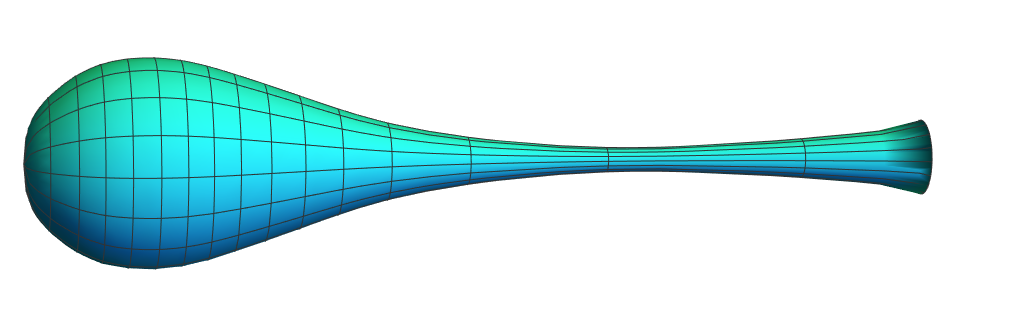} \includegraphics[width=0.36\textwidth]{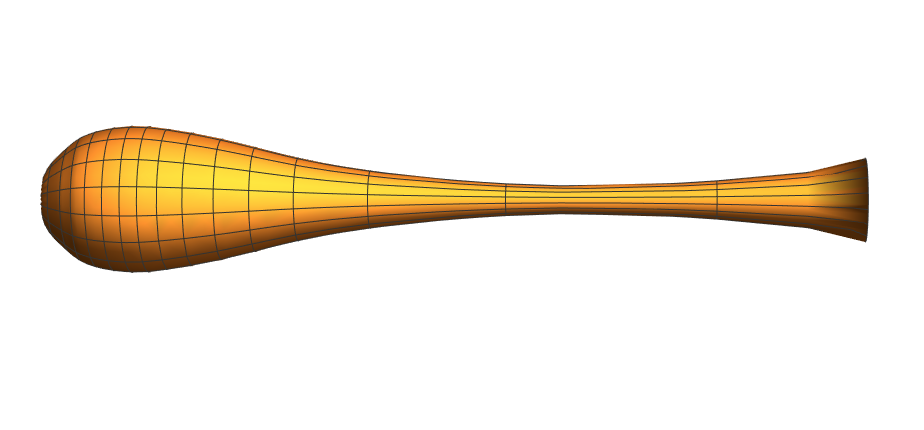}
 \caption{Embedding in Euclidean space of the boundary (left) and the horizon (right) of a ``bottle shaped" black hole in region D4 with $\mu =3.8$, $\nu=3.895$, close to the D2 border at $\nu = 2\sqrt{\mu} \approx 3.899$. \label{Fig:spikeD4}}}
 \end{center}
\end{figure}

In most cases, the boundary geometry resembles the horizon geometry. This is shown in 
Fig. \ref{Fig5} where we have embedded both the boundary geometry and horizon geometry in three-dimensional hyperbolic  space to make them easy to visualize.
Comparing the boundary and horizon geometries, one can show the following:
\begin{itemize}
\item A globule on the horizon is present only if there is a corresponding globule on the boundary, however the converse is not true: there are configurations where the boundary has a globule and the horizon only has a ``mountain-like" deformation near the axis.
\item The scalar curvature at the tip of the globule is always larger on the horizon than on the boundary. 
\item Near the boundary with D2, where the horizon develops a long spike, the boundary geometry also develops a long spike (see Fig. \ref{Fig:spikeD4}). This is because $G(x)$ is developing a double zero at $x_-$.
\end{itemize}

So, as expected,  the deformations of the horizon geometry are caused by corresponding deformations of the boundary geometry. Even dramatic changes to the horizon (such as long spikes) can be traced to corresponding dramatic changes to the boundary. 

\begin{figure}[H]
\begin{center}
 {\includegraphics[width=.44\textwidth]{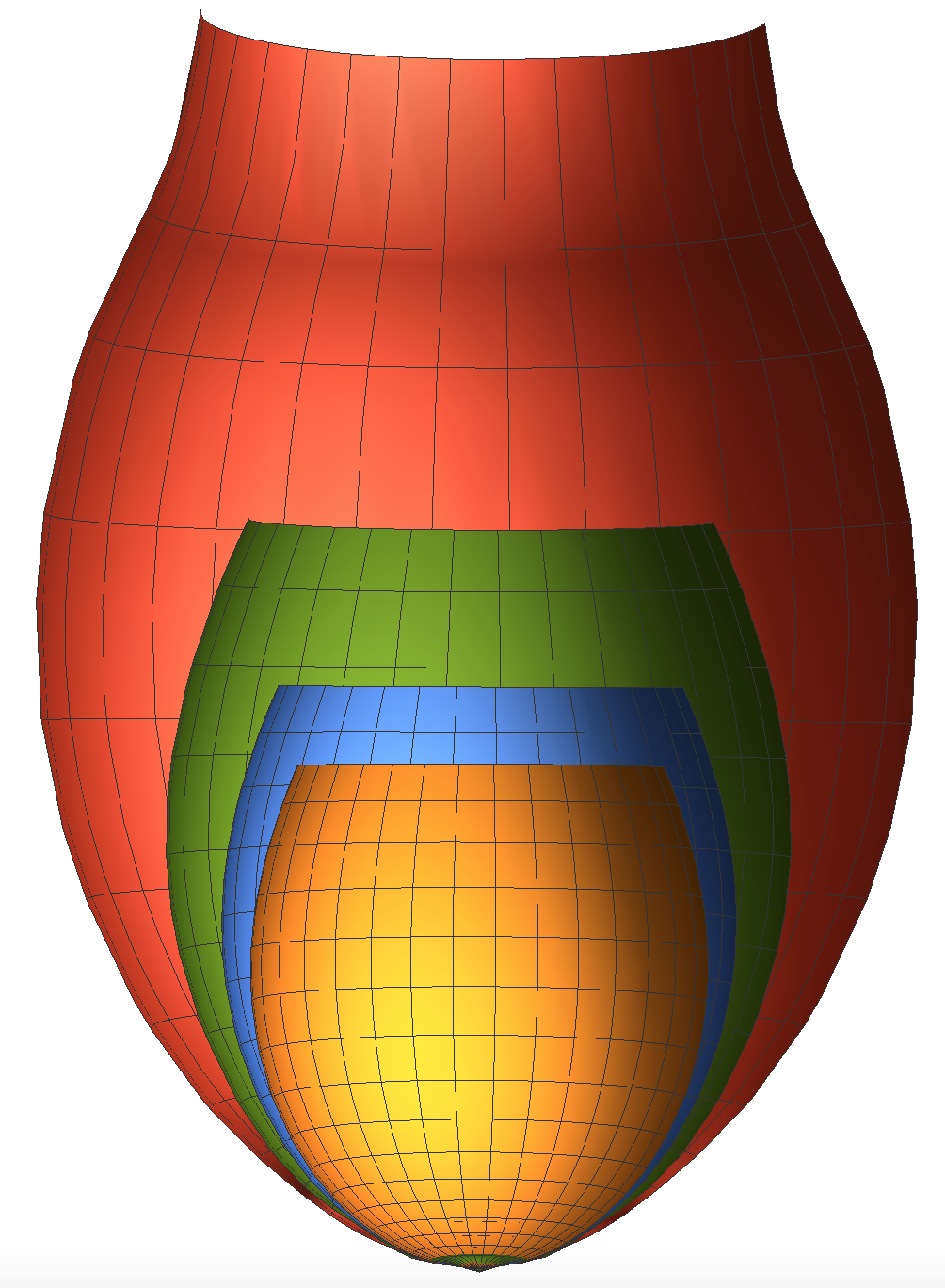} \includegraphics[width=.36\textwidth]{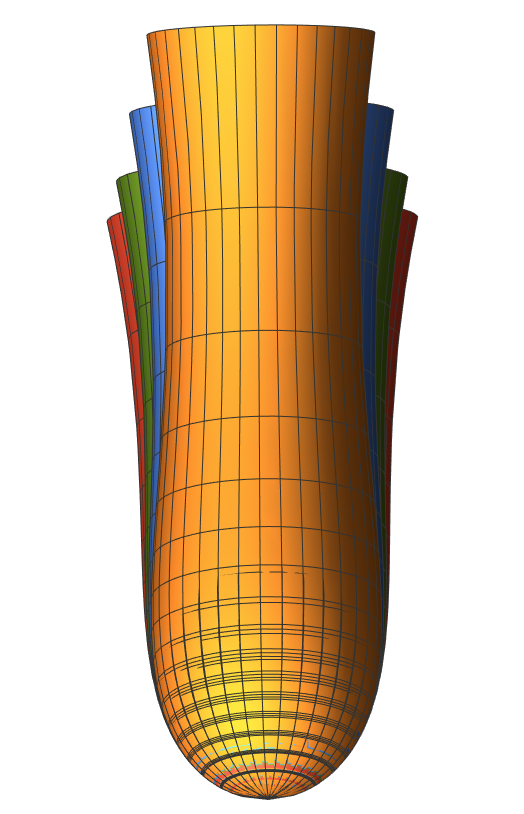}  
\caption{\label{Fig18} The left figure shows the tip of the boundary geometries of region D4 embedded in Euclidean space for a fixed value of $\mu =13$ for the following values: $\nu = 6.7$ (yellow), $\nu = 6.45$ (blue), $\nu = 6.2 $ (green), $\nu = 5.98$ (red). The red figure is close to the D3-D4 boundary, which is located at $\nu= \mu -2\sqrt{\mu} \simeq 5.79$. The right figure shows the tip of the associated horizon geometry. One can see that, while the change in the boundary is large approaching the boundary D3-D4, the associated change in the horizon geometry is very small. The geometries all approach a unit hyperboloid at infinity.}}
 \end{center}
\end{figure}

However, there is a region of parameters where the boundary undergoes a dramatic change and the horizon does not respond much at all.  This occurs at the D4/D3 interface where $F(x)$ develops a double root at $x= y_+ = -1 + 1/\sqrt \mu$. We have seen that the horizon geometry changes smoothly across this interface. However from the form of the boundary metric \eqref{withfactor}, it is clear that this causes a huge change on the boundary. The region of space around $x = y_+$ is stretched enormously. In fact, any point with $x<y_+$ becomes infinitely far away from any point with  $x > y_+$ at the D4/D3 interface. The boundary geometry spits into two disconnected geometries. When a globule is present in the boundary metric, it becomes infinitely big at this interface (see Fig. \ref{Fig18}). Note that the curvature of the horizon on the axis in Fig. \ref{Fig18} does not change as we vary $\nu$, since it only depends on $\mu$ \eqref{horcurv}.

   This is surprising, as a large change in the boundary should lead to a corresponding  large effect on the horizon.  More importantly, it also raises a problem with gauge/gravity duality.  The dual CFT is presumably in a thermal state, so the entropy should be proportional to the volume of space. (Since $g_{tt}=-1$ there is no redshift correction.) As one approaches the D4/D3 boundary, the entropy computed in the boundary increases without bound, but the entropy computed in the bulk does not.  To make this statement precise, we need to regulate the infinite volume associated with the asymptotic hyperbolic region. This can be done as follows. Fix a $\phi$ circle in the asymptotic region with circumference $C_\star$. From \eqref{withfactor}, the area of the boundary up to this circle is
   \be\label{regarea} 
   A_\partial^{C_\star} = \frac{(1+\nu)^2}{4} \Delta \phi \int_{-1}^{\epsilon} \frac{dx}{[-F(x)]^{3/2}}
  \ee
  where $x =\epsilon$ labels the circle with circumference $C_\star$.
  This regulates the divergence associated with the fact that $F$ vanishes at $x=0$.
  We now subtract the area of a unit radius two-dimensional hyperbolic space, up to the same value of $C_\star$ (we also choose $\phi$ in this auxiliary space to have the same period as the solutions we constructed). This difference has a finite limit as we take $C_\star \to \infty$. We denote this limit  by $\mathcal{I}_{\partial}$. We then repeat the same procedure for the horizon area (computed from \eqref{metric_horizon}) and denote this quantity by $\mathcal{I}_{\mathcal{H}}$.  This procedure removes the  divergence due to the asymptotics. The spirit of this renormalisation is similar to holographic renormalisation:  for large $C_\star $, the area of the hyperbolic space is simply $C_\star - \Delta \phi$,  so the divergent term  we are subtracting  is the only local counter-term one can write in one dimension. It is now clear from \eqref{regarea} that if $F(x)$ develops a zero at any nonzero value of $x$ (as it does  at the D4/D3 boundary) $\mathcal{I}_{\partial}$ will diverge while $\mathcal{I}_{\mathcal{H}}$ does not. This suggests that there is another branch of bulk solutions with the same boundary geometry for which the horizon area does grow without bound. We will construct such solutions numerically in the next section.
  
\vspace{10mm}

Before we proceed to that construction, we would like to determine the holographic stress energy tensor associated with such boundary metric. This is best done by using Fefferman-Graham coordinates, where the bulk metric takes the following schematic form \cite{Fefferman:2007rka}
\be
\mathrm{d}s^2 = \frac{1}{z^2} \Big[\mathrm{d}s^2_{\partial}+z^2\,\mathrm{d}s^2_{2}+z^3\,\mathrm{d}s^2_{3}+\mathcal{O}(z^4)+\mathrm{d}z^2\Big]\,,
\label{boundary_m}
\ee
and the holographic stress energy tensor is built using the relation \cite{deHaro:2000vlm}
\be
\langle T_{\alpha\beta}\rangle\mathrm{d}x^\alpha\mathrm{d}x^\beta=\frac{3}{16\,\pi\,G_4}\mathrm{d}s^2_3\,,
\ee
where here henceforth lower Greek indices run over the boundary coordinates. In order to find the relation between $(t,x,y)$ and $(\tilde{t},r,z)$ we equate Eq.~(\ref{boundary_m}) with our line element (\ref{C-metric}). This allows us to express $x$ and $y$ as a power series in $z$, by demanding that the boundary metric takes the form given in Eq.~(\ref{eq:boundarymetric}), $g_{z\alpha}=0$ and $g_{zz}=1/z^2$. The first few terms in such expansion read
\begin{subequations}
\begin{align}
&t=\frac{2}{1+\nu}\tilde{t}\,,
\\
& x = w^2(2-w^2)-1+\alpha_0(w)\,z+\mathcal{O}(z^2)\,,
\\
& y = w^2(2-w^2)-1+\gamma_0(w)\,z+\mathcal{O}(z^2)\,,
\end{align}
\label{eq:bulktoboundary}
\end{subequations}
where
\begin{subequations}
\begin{align}
&\alpha_0(w)=-\frac{2 w^2 \left(1-w^2\right) \left(2-w^2\right) \sqrt{F_w(w)} G_w(w)}{1+\nu }\,,
\\
&\gamma_0(w)=\frac{2 \left(1-w^2\right)^3 F_w(w){}^{3/2}}{1+\nu}\,,
\\
&F_w(w)=1- (\mu -\nu )\left(2-w^2\right) w^2+\mu  \left(2-w^2\right)^2 w^4\,,
\\
& G_w(w)=1-\nu  \left(1-w^2\right)^2+\mu  \left(1-w^2\right)^4\,.
\end{align}
\label{eq:def}
\end{subequations}
We have introduced a new boundary radial coordinate  $w\in[0,1]$ that will be convenient for the numerical work in the next section. Its relation to $r$ will be given momentarily. In the $w$ coordinate, the boundary metric reads
\be
\mathrm{d}s^2_{\partial} =-\mathrm{d}\tilde{t}^2+\frac{(1+\nu )^2 }{4}\Bigg[\frac{16\,\mathrm{d}w^2}{\left(1-w^2\right)^2 \left(2-w^2\right) F_w(w){}^2 G_w(w)}+\frac{w^2 \left(2-w^2\right) G_w(w)}{\left(1-w^2\right)^2\,F_w(w)}\mathrm{d}\phi^2 \Bigg]\,,
\ee
which allows us to read the proper radius $r$ in \eqref{eq:boundarymetric} as
\be
\int_{0}^{w}\frac{2 (1+\nu )}{\left(1-\tilde{w}^2\right) \sqrt{2-\tilde{w}^2} F_w(\tilde{w}) \sqrt{G_w(\tilde{w})}}\mathrm{d}\tilde{w}=r\,.
\ee
Although we have written our boundary metric in the $w$ coordinates, it is trivial to return to our original $x$ coordinates via Eqs.~(\ref{eq:bulktoboundary}) and it reduces to \eqref{withfactor}.

The final expression for the holographic stress energy tensor is rather lengthy. However we note that since our boundary metric is static and axisymmetric, the holographic stress energy tensor has a single independent component\footnote{Recall that the holographic stress energy tensor is necessarily traceless and covariantly conserved.}, which we can choose to be the energy density $\rho\equiv - \langle T^{\tilde{t}}_{\phantom{\tilde{t}}\tilde{t}}\rangle$. In $w$ coordinates, it reads
\be
\rho = \frac{\left(1-w^2\right)^3 \mu}{2 \pi  G (1+\nu )^3}  F_w(w){}^{3/2}\left[2-3 w^2 \left(2-w^2\right) G_w(w)\right]\,.
\label{eq:stress_energy}
\ee
The total energy can then be readily computed via a simple integration
\be
E = -\frac{\mu  [\mu -2 (\nu +1)]}{8 G ( 1+\nu) (1+\mu -\nu )}\,,
\ee
where we used
\be
\Delta \phi = \frac{4\pi}{|1+\mu-\nu|}\,.
\label{eq:peridocitiy}
\ee
Note that even though the boundary has infinite area, the total energy is finite. This is because the solutions all have $T=1/2\pi$ and a homogeneous hyperbolic black hole with this temperature is locally AdS and has zero energy. The localized deformation adds only a finite amount of energy. Near the D4/D3 boundary, $\rho\rightarrow 0$ in the region of growing area on the boundary, so $E$ still remains bounded.
 \begin{figure}[H]
  \begin{center}
\includegraphics[width=0.5\textwidth]{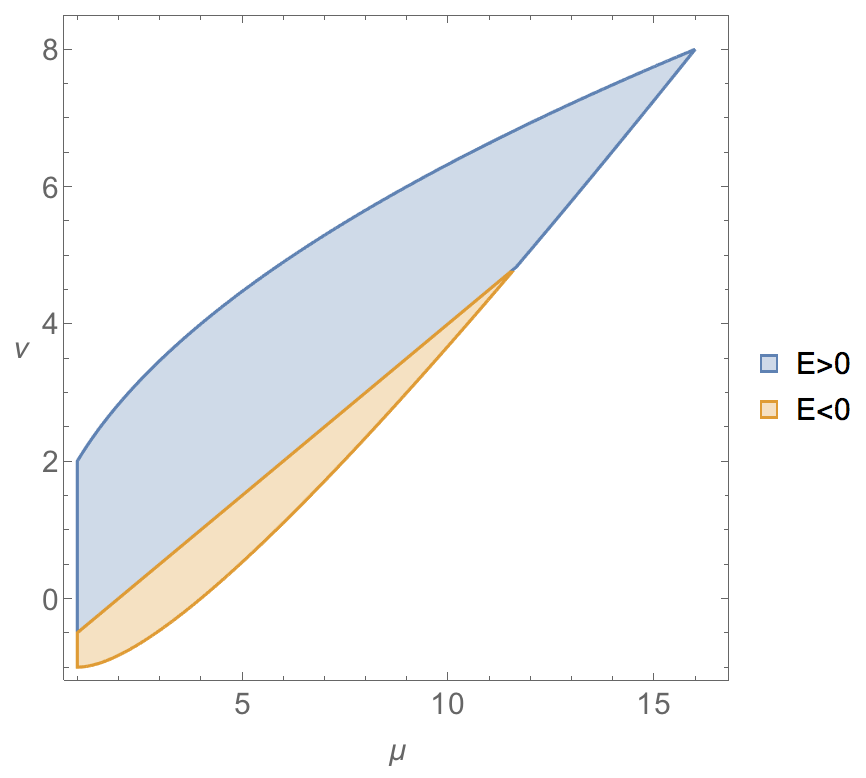}
\caption{Moduli space of region D4 where the energy can become negative.}
\label{fig:energy}
\end{center}
\end{figure}
It is interesting to note that the total energy can be negative in the region $\mu-2\sqrt{\mu}<\nu<\mu/2-1$ with $\mu\in(1,6+4\sqrt{2})$, but that this does not happen along the whole D3/D4 boundary. In Fig.~\ref{fig:energy} we show the moduli space of region D4 where the energy can be negative. It is not unusual for hyperbolic black holes to have $E < 0$, since even homogeneous black holes can have negative energy density.

\subsection{New solutions in region D4 \label{nregD4}}
\subsubsection{\emph{Ansatz}}
We are going to construct new solutions with boundary metrics as in D4 using the DeTurck trick, which was first introduced in \cite{Headrick:2009pv} and reviewed in \cite{Wiseman:2011by,Dias:2015nua}. The idea is to deform the Einstein equation into the so called Einstein-DeTurck equation
\begin{equation}
R_{ab}+{3}g_{ab}-\nabla_{(a}\xi_{b)}=0\,,
\label{eq:deturck}
\end{equation}
where $\xi^a = g^{bc}\left[\Gamma^{a}_{bc}(g)-\Gamma^{a}_{bc}(\bar{g})\right]$, $\Gamma(\mathfrak{g})$ is the Christoffel connection associated with a metric $\mathfrak{g}$ and $\bar{g}$ is the so called reference metric. The reference metric $\bar{g}$ essentially determines our gauge choice and, apart from sharing the same causal structure as the solution $g$ we wish to find, is otherwise arbitrary. Solutions of the Einstein-DeTurck equation will only coincide with solutions of the Einstein equation if $\xi=0$. For the system at hand, it has been shown in \cite{Figueras:2011va} that solutions with $\xi\neq0$ do not exist. This result follows from the fact that the solution we seek is static, and from the contracted Bianchi identity\footnote{Note however that in \cite{Figueras:2016nmo} a novel proof for the non-existence of solutions with $\xi\neq0$ has appeared where only stationarity and $t-\phi$ symmetry are required.}. The advantage of solving Eq.~(\ref{eq:deturck}) instead of solving the original Einstein equations is immense, namely Eq.~(\ref{eq:deturck}) is manifestly elliptic, and can be readily solved using a standard Newton-Raphson relaxation routine.

Our metric \emph{Ansatz} reads
\begin{multline}
\mathrm{d}s^2 = \frac{1}{(1-\tilde{y}^2)^2}\Bigg\{-\tilde{y}^2\,G_{\tilde{y}}(\tilde{y})\,q_1(w,\tilde{y})\,\mathrm{d}\tilde{t}^2+\frac{4\,y_+^2\,q_2(w,\tilde{y})}{G_{\tilde{y}}(\tilde{y})}\left(\mathrm{d}\tilde{y}+\frac{q_3(w,\tilde{y})}{1-w^2}\mathrm{d}w\right)^2+\\
\frac{y_+^2(1+\nu )^2 }{4}\Bigg[\frac{16\,q_4(w,\tilde{y})\,\mathrm{d}w^2}{\left(1-w^2\right)^2 \left(2-w^2\right) F_w(w)^2 G_w(w)}+\frac{w^2 \left(2-w^2\right) G_w(w)\,q_5(w,\tilde{y})}{\left(1-w^2\right)^2\,F_w(w)}\mathrm{d}\phi^2 \Bigg]\Bigg\}
\label{eq:ansatz}
\end{multline}
where $G_w(w)$ and $F_w(w)$ are given as in Eqs.~(\ref{eq:def}) and
\begin{equation}
G_{\tilde{y}}(\tilde{y})=\left(3-3 \tilde{y}^2+\tilde{y}^4\right) y_+^2-\left(1-\tilde{y}^2\right)^2\,.
\end{equation}
Some remarks are in order regarding the \emph{Ansatz} above. First, our domain of integration is a square $(w,\tilde{y})\in[0,1]\times[0,1]$. Second, for our reference metric we will take the line element above with $q_1=q_2=q_4=q_5=1$ and $q_3=0$. Finally, our solution is fully determined once the five functions of two variables $q_i(w,\tilde{y})$ with $i=\{1,2,3,4,5\}$ are found.

We now turn our attention to the boundary conditions imposed at the edge of our computational domain. The conformal boundary is located at $\tilde{y}=1$, where we demand that $q_1=q_2=q_4=q_5=1$ and $q_3=0$. The induced metric on such a hyperslice, to leading order, in $(1-\tilde{y})$ reduces to
\begin{multline}
\left.\mathrm{d}s^2\right|_{\tilde{y}\sim 1} = \frac{1}{4(1-\tilde{y})^2}\Bigg\{-y_+^2\mathrm{d}\tilde{t}^2+4\mathrm{d}\tilde{y}^2+\\
\frac{y_+^2(1+\nu )^2 }{4}\Bigg[\frac{16\mathrm{d}w^2}{\left(1-w^2\right)^2 \left(2-w^2\right) F_w(w)^2 G_w(w)}+\frac{w^2 \left(2-w^2\right) G_w(w)}{\left(1-w^2\right)^2\,F_w(w)}\mathrm{d}\phi^2 \Bigg]\Bigg\}\,,
\end{multline}
which under the coordinate transformation
\begin{equation}
\tilde{y}=1-\frac{y_+}{2}z\,,
\end{equation}
brings the boundary metric to the relevant conformal frame.

The horizon is the null hypersurface $\tilde{y}=0$, and our choice of reference metric dictates that $q_1(w,0)=q_2(w,0)$. This, in turn, implies that the temperature associated with the line element above is given by
\begin{equation}
T=\frac{3 y_+^2-1}{4 \pi  y_+}\,.
\label{eq:temp}
\end{equation}
We see that $y_+$ controls the temperature of the black hole horizon, with Eq.~(\ref{eq:tempD4}) being recovered for $y_+=1$. The remaining boundary conditions are simply of the Neumann type: $\partial_{\tilde{y}} q_i|_{\tilde{y}=0}=0$ for $i=\{1,2,4,5\}$ and $q_3|_{\tilde{y}=0}=0$. Furthermore, extremality can be achieved if we take $y_+=1/\sqrt{3}$.

At $w=0$, the circle parametrised by $\phi$ shrinks to zero size, so this marks the axis of rotation. This in turn implies the following boundary conditions $q_4(0,\tilde{y})=q_5(0,\tilde{y})$, $q_3(0,\tilde{y})=0$ and Neumann for the remaining variables: $\partial_w q_i|_{w=0}=0$ for $i=\{1,2,4,5\}$. The first of these conditions ensures that the periodicity of $\phi$ is given by Eq.~(\ref{eq:peridocitiy}).

Finally, we come to $w=1$, which locates the region infinitely far away from the axis of rotation. Here, the bulk metric should approach a solution of the Einstein equation with a negative cosmological constant, at temperature (\ref{eq:temp}) and with a horizon with hyperbolic geometry. One such solution is the hyperbolic black hole which is recovered if we set $q_1=q_2=q_4=q_5=1$, $q_3=0$, identify
\begin{equation}
r=\frac{r_+}{1-\tilde{y}^2}\,,\quad y_+\equiv r_+\,,\quad M_{\mathrm{Hyper}}\equiv\frac{r_+}{2}\left(r_+^2-1\right)
\end{equation}
and expand the line element (\ref{eq:ansatz}) close to $w=1$.

\subsubsection{Numerical results}
\begin{figure}[H]
\begin{center}
\includegraphics[width=.62\textwidth]{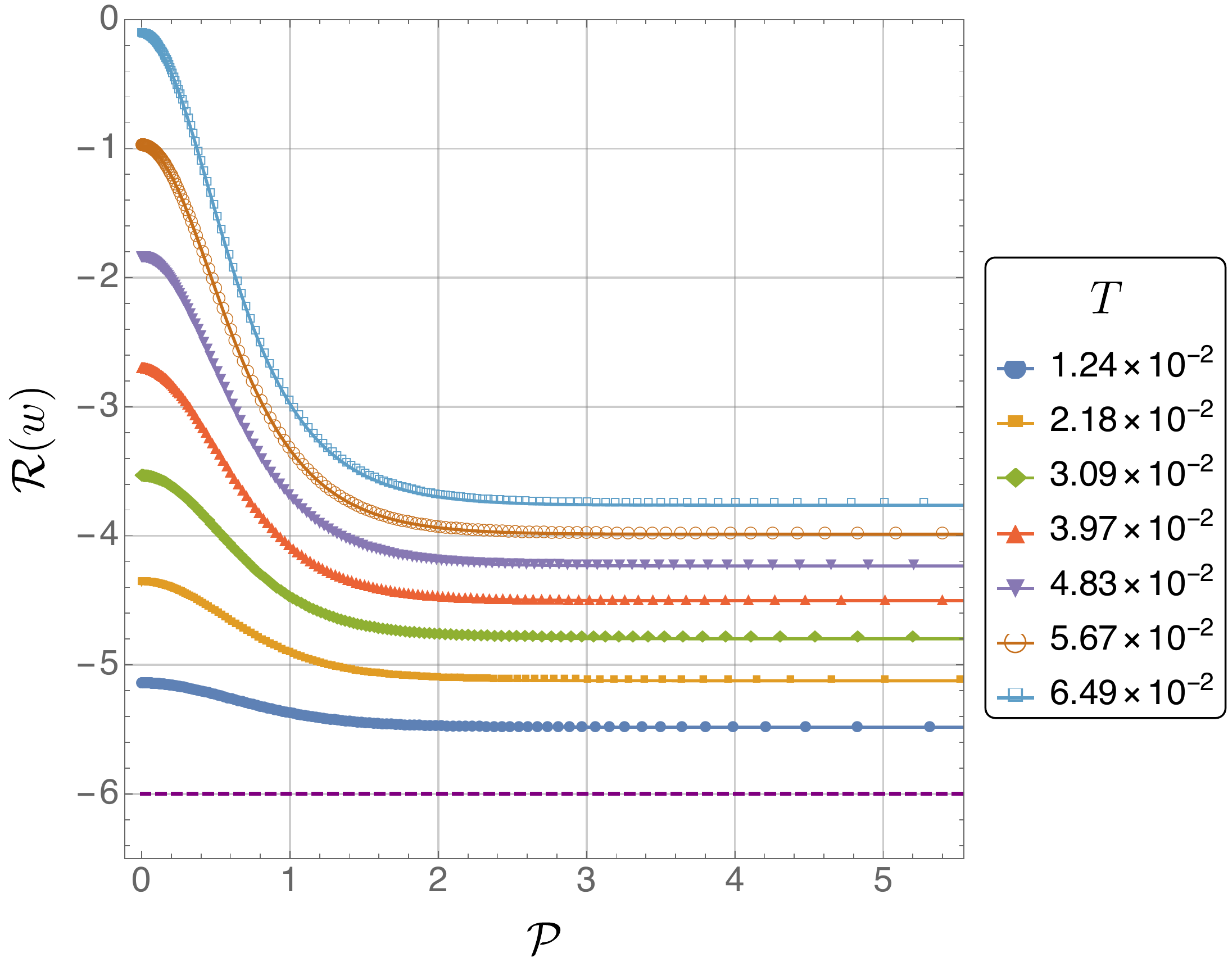}
 \caption{\label{fig:curvature}Curvature of the induced horizon geometry as a function of the proper distance, along the horizon, from the axis of rotation. This plot was generated with $\mu=3$ and $\nu=8\sqrt{3}/5$, and the range of temperatures is indicated in the legend above. The horizontal purple dashed curve indicates the curvature of an extremal uniform hyperbolic black hole.}
 \end{center}
\end{figure}

We will first describe what happens if we take the black hole solutions described in \ref{RegD4}, but change their asymptotic temperature. We have constructed a number of solutions for different values of $\mu$ and $\nu$ lying in D4, and the results appear universal. In Fig.~\ref{fig:curvature} we show the induced horizon curvature as a function of the proper distance, along the horizon, from the axis of rotation. To generate this figure we have used $\mu=3$ and $\nu=8\sqrt{3}/5$. As the system is cooled down, the horizon becomes less floppy and starts approaching a geometry with constant curvature. We believe that if we were able to reach zero temperature, the curvature would become exactly $-6$, corresponding to a uniform extremal hyperbolic black hole in four spacetime dimensions.

\begin{figure}[H]
\begin{center}
\includegraphics[width=.45\textwidth]{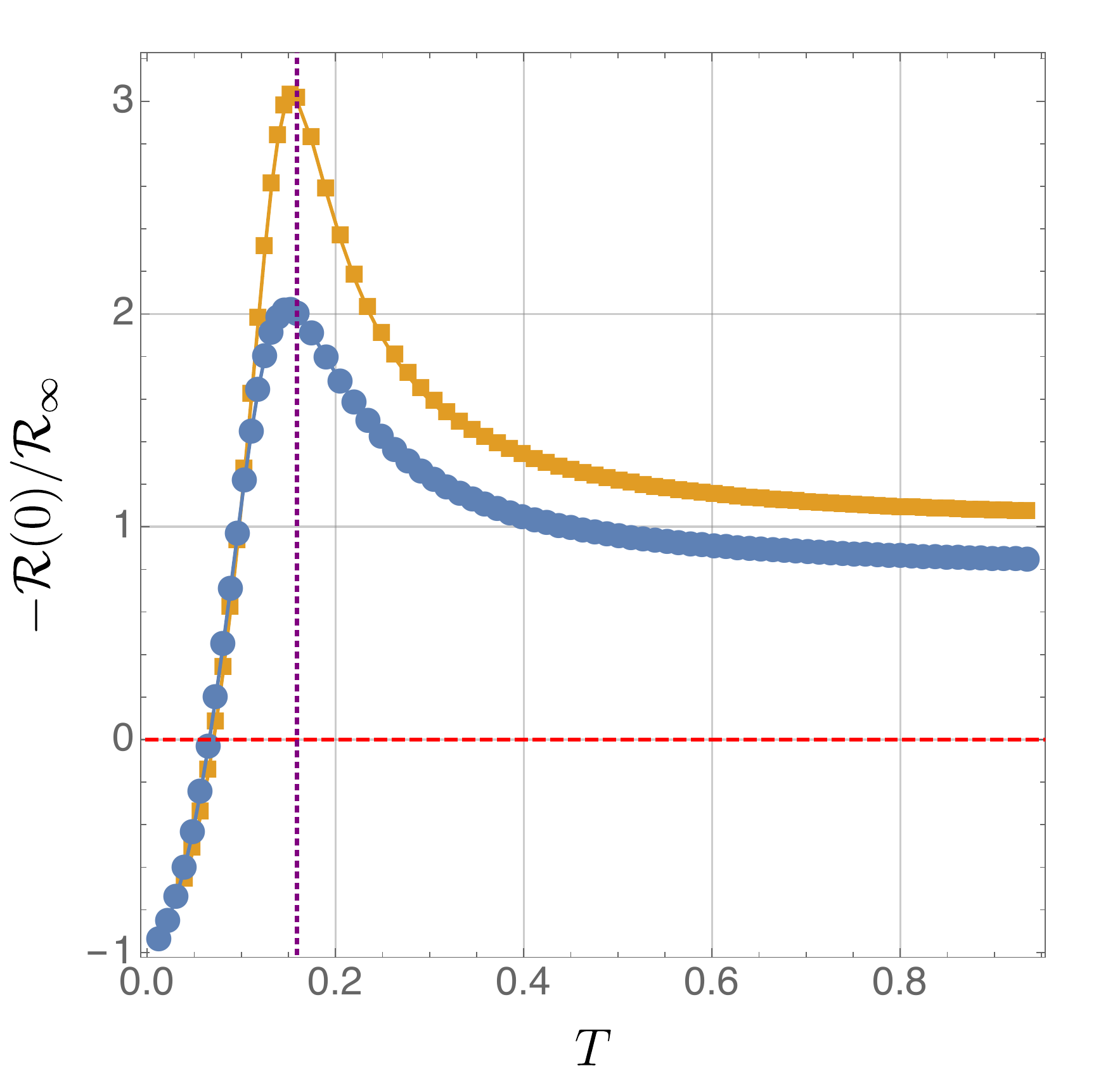}
 \caption{\label{fig:curvature2}Induced horizon curvature at the axis of rotation (measured in units  of curvature corresponding to a uniform hyperbolic black hole at the same temperature, $\mathcal{R}_{\infty}=-2/y_+^2$) as a function of temperature. The orange squares have $\mu=10$, $\nu=4$ and the blue disks have $\mu=3$ and $\nu=8\sqrt{3}/5$. The purple dotted vertical line indicates $T=1/(2\pi)$ and the red horizon line indicates $\mathcal{R}=0$.}
 \end{center}
\end{figure}

We have also followed the induced horizon curvature at the axis of rotation (measured in units  of curvature corresponding to a uniform hyperbolic black hole at the same temperature, $\mathcal{R}_{\infty}=-2/y_+^2$) as a function of the black hole temperature. This is displayed in Fig.~\ref{fig:curvature2} for two different values of $(\mu,\nu)$: the orange squares have $(\mu,\nu)=(10,4)$ and the blue disks have $(\mu,\nu)=(3,8\sqrt{3}/5)$. The temperature corresponding to the analytic solutions of section \ref{RegD4} is indicated as the purple dotted vertical line with $T=1/(2\pi)$. It is interesting to note that the maximum deformation seems to occur close to $T=1/(2\pi)$.

At high temperatures we expect the fluid/gravity approximation of \cite{Bhattacharyya:2008jc,Bhattacharyya:2008xc} to be exact. If this is the case, the horizon curvature should approach the boundary curvature. This is indeed the case, as we can confirm in Fig.~\ref{fig:curvature2}: at large $T$ the curvature seems to saturate at the value predicted by fluid/gravity correspondence given by
\begin{equation}
-\frac{\mathcal{R}(0)}{\mathcal{R}_{\infty}}=\frac{2 \mu  (3+\nu)-\mu ^2-(1+\nu)^2}{(1+\nu)^2}\,.
\end{equation}

\begin{figure}[H]
\begin{center}
\includegraphics[width=.6\textwidth]{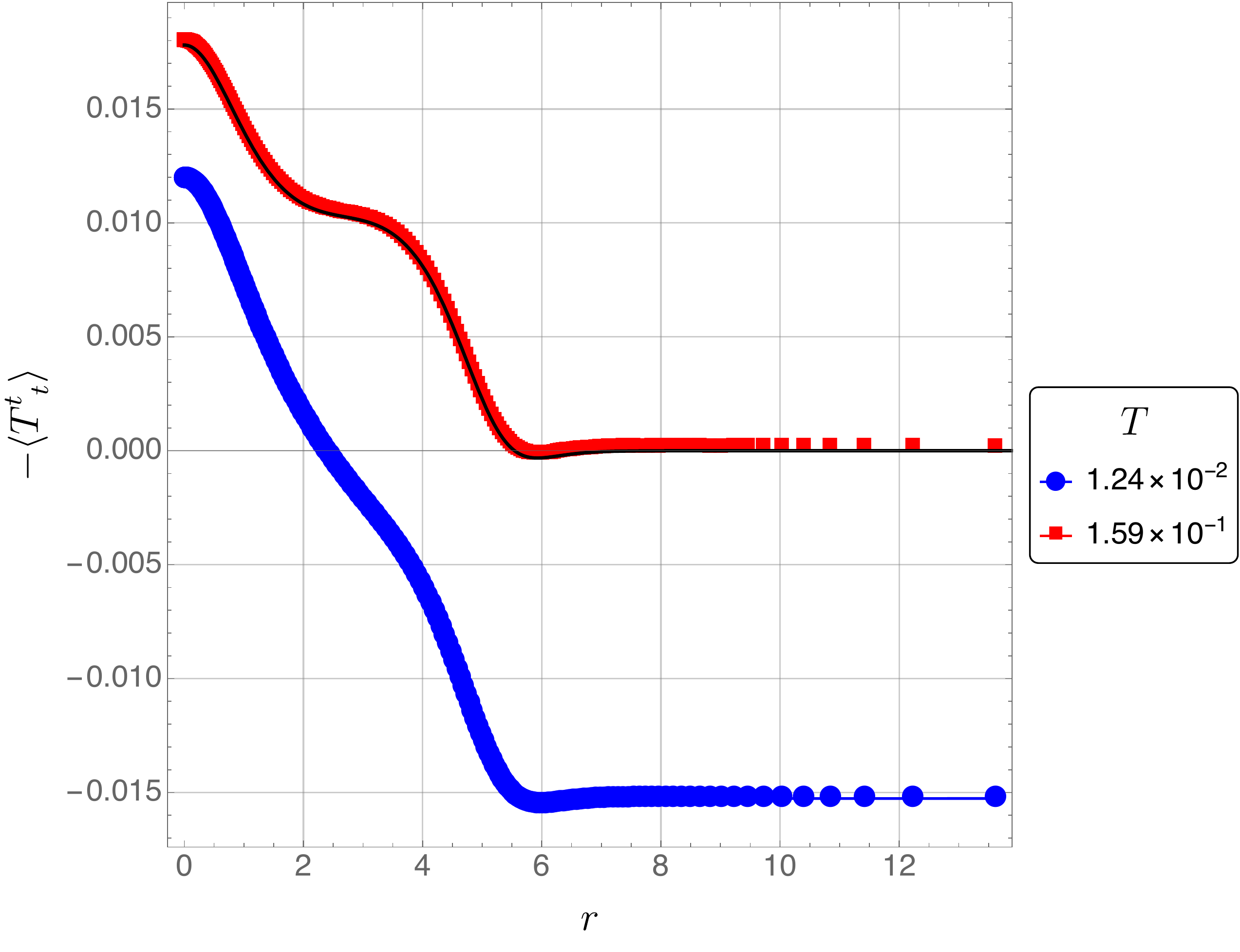}
 \caption{\label{fig:stress}Energy density as a function of the boundary proper distance $r$, computed with $\mu=3$, $\nu=8\sqrt{3}/5$ (same parameters as in Fig.~\ref{fig:curvature}). The solid black line corresponds to the analytic solution Eq.~(\ref{eq:stress_energy}), and the red squares and blue disks to numerical data. The temperatures used are indicated in the legend above.}
 \end{center}
\end{figure}

We have also computed the energy density as a function of the boundary proper distance $r$ for different temperatures (see Fig.~\ref{fig:stress} for $T=1/(2\pi)$ and a lower temperature,  using the same parameters as those used in Fig.~\ref{fig:curvature}). For $T=1/(2\pi)$ this can also be extracted analytically, so we added this curve to our plot (corresponding to the solid black line). The agreement between our numerics (represented as the red squares) and the analytic expression Eq.~(\ref{eq:stress_energy}) is reassuring. We see that as the temperature is lowered, the energy density decreases eventually making the total energy negative. 

In addition to generalising the solution of section \ref{RegD4} to several asymptotic temperatures, we have also found a new phase, corresponding to a new black hole solution that is not captured by the C-metric. Furthermore, this novel black hole phase resolves the puzzle raised in section 4.1 since the black hole horizon responds to the large deformations in the boundary metric when approaching the D3/D4 boundary. In order to quantify this, let us recall the regularization that was described around Eq. \eqref{regarea}.

We take a cut-off great circle of circumference $C_\star$, and compute the area of the boundary metric minus that of a unit radius two-dimensional hyperbolic space, up to the same value of $C_\star$ (we also choose $\phi$ in this auxiliary space to have the same period as the solutions we constructed). This difference has a finite limit as we take $C_\star \to \infty$. We denote this limit  by $\mathcal{I}_{\partial}$. We then repeat the same procedure for the horizon area and denote this quantity by $\mathcal{I}_{\mathcal{H}}$. 

\begin{figure}[H]
\begin{center}
\includegraphics[width=.6\textwidth]{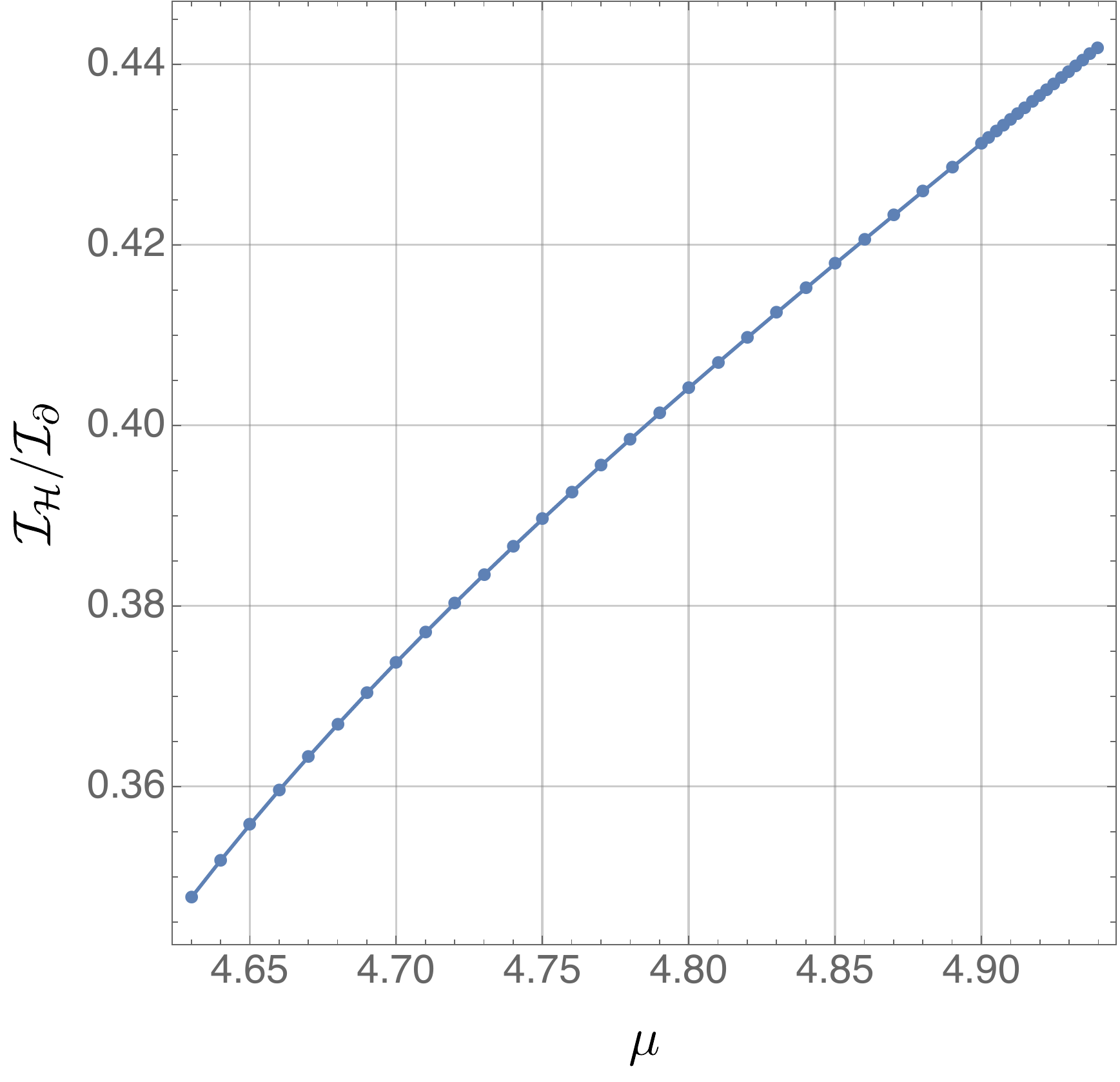}
 \caption{\label{fig:finite}A comparison of the (regulated) area of the horizon and boundary for the new solutions as a function of $\mu$ for $T=1/(2\pi)$ and $\nu=1/2$. The fact that the ratio does not vanish near the boundary with D3 ($\mu \approx 4.949$) shows the horizon area is increasing at the same rate.}
 \end{center}
\end{figure}

Any divergence that is left has to do with the bulging of the horizon or boundary. For the new solutions we constructed, the ratio $\mathcal{I}_{\mathcal{H}}/\mathcal{I}_{\partial}$ is non-zero (but finite) as we approach the D3/D4 boundary (see Fig.~\ref{fig:finite} which was generated for $\nu=1/2$ and $T=1/(2\pi)$), while for the analytic solutions of section \ref{RegD4} the regulated area of the black hole horizon is identically zero\footnote{To be precise, one can show that $\mathcal{I}_{\mathcal{H}}\propto \mathcal{\mathcal{O}}(C_{\star}^{-1})$.}.

\begin{figure}[H]
\begin{center}
\includegraphics[width=.52\textwidth]{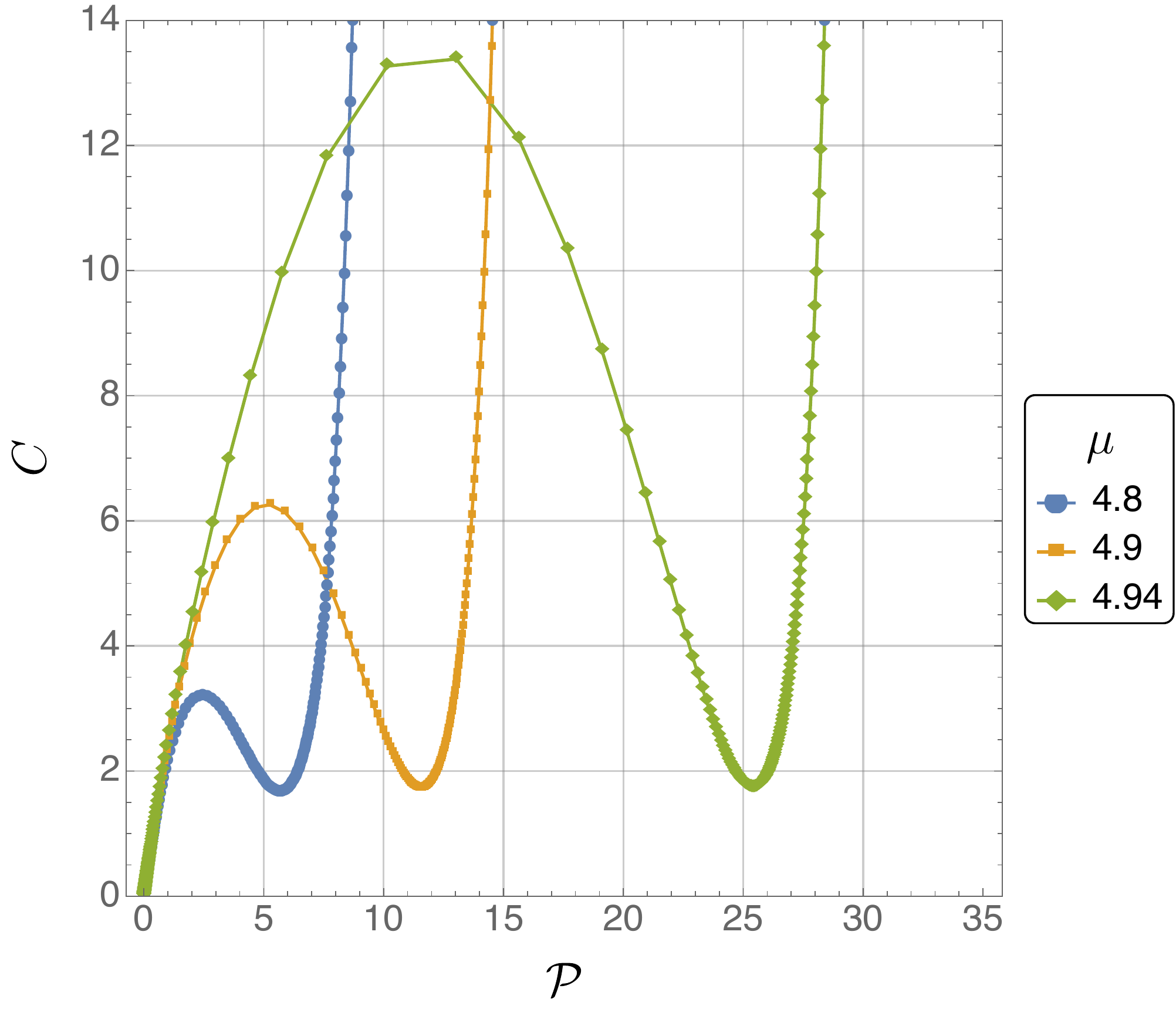}
 \caption{\label{fig:horizonnew}Circumference of the $\phi$ circles as a function of the proper distance $\mathcal{P}$ for constant $\nu = 1/2$, and varying $\mu$ towards the boundary with D3 ($\mu \approx 4.949$).}
 \end{center}
\end{figure}

In order to visualize the new horizons, we plot in Fig.~\ref{fig:horizonnew} the circumference of the $\phi$ circles, $C$, as a function of the proper distance $\mathcal{P}$ for constant $\nu = 1/2$, and varying $\mu$ towards the D3/D4 boundary. As expected, the closer to the boundary (corresponding to $\mu \approx 4.949$), the more the horizon geometry bulges.

\begin{figure}[H]
\begin{center}
\includegraphics[width=.62\textwidth]{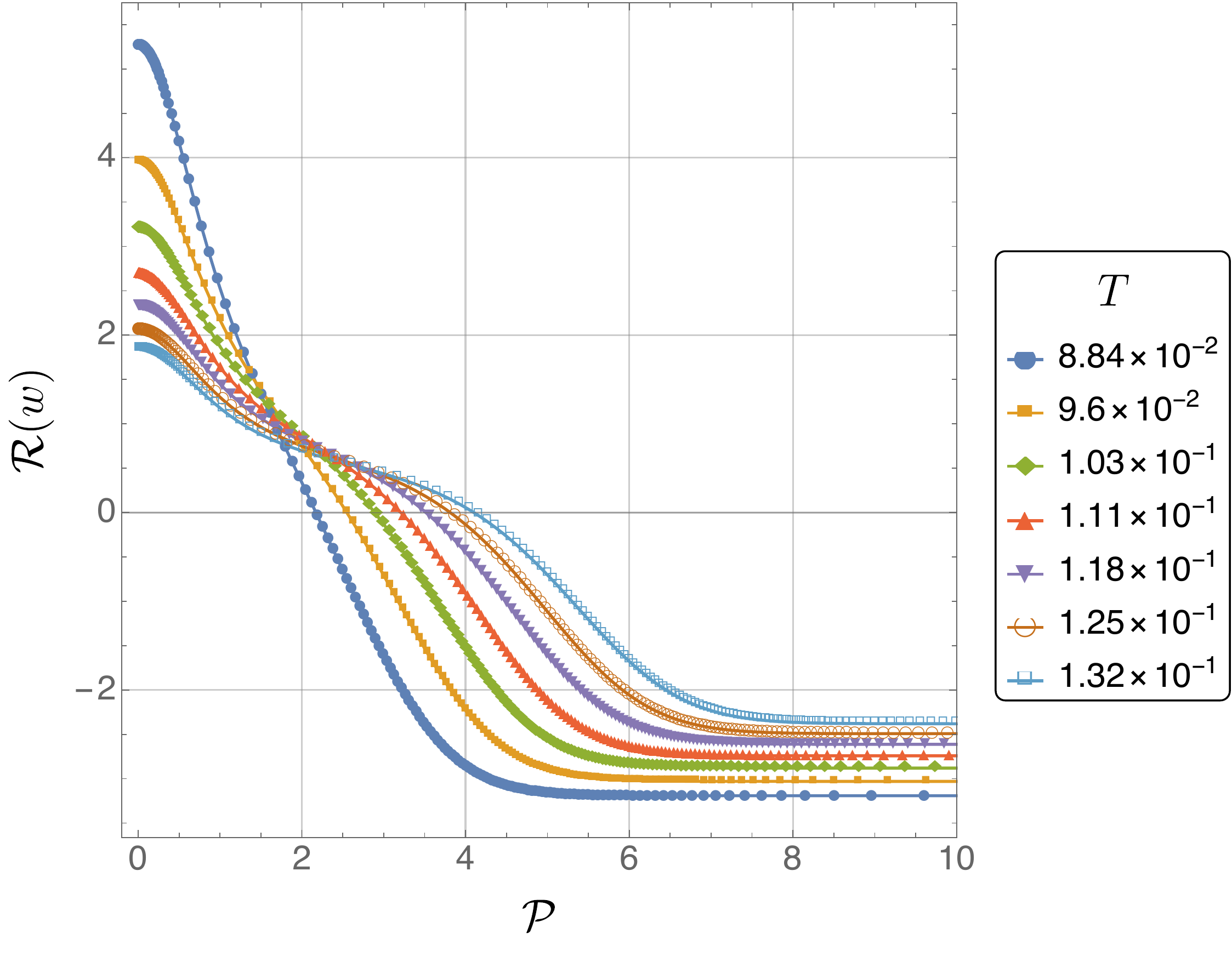}
 \caption{\label{fig:bulge}Curvature of the induced horizon geometry of the novel black hole phase as a function of the proper distance, along the horizon, from the axis of rotation. This plot was generated with $\mu=4$ and $\nu=3$, and the range of temperatures are indicated in the legend.}
 \end{center}
\end{figure}

This new phase also has a very different behaviour when we vary the temperature. We start by plotting the induced horizon curvature as a function of the proper distance to the axis of rotation (see Fig.~(\ref{fig:bulge})). As the system cools down, the induced horizon curvature measured at the axis of rotation increases, apparently without bound.

\begin{figure}[H]
\begin{center}
\includegraphics[width=.6\textwidth]{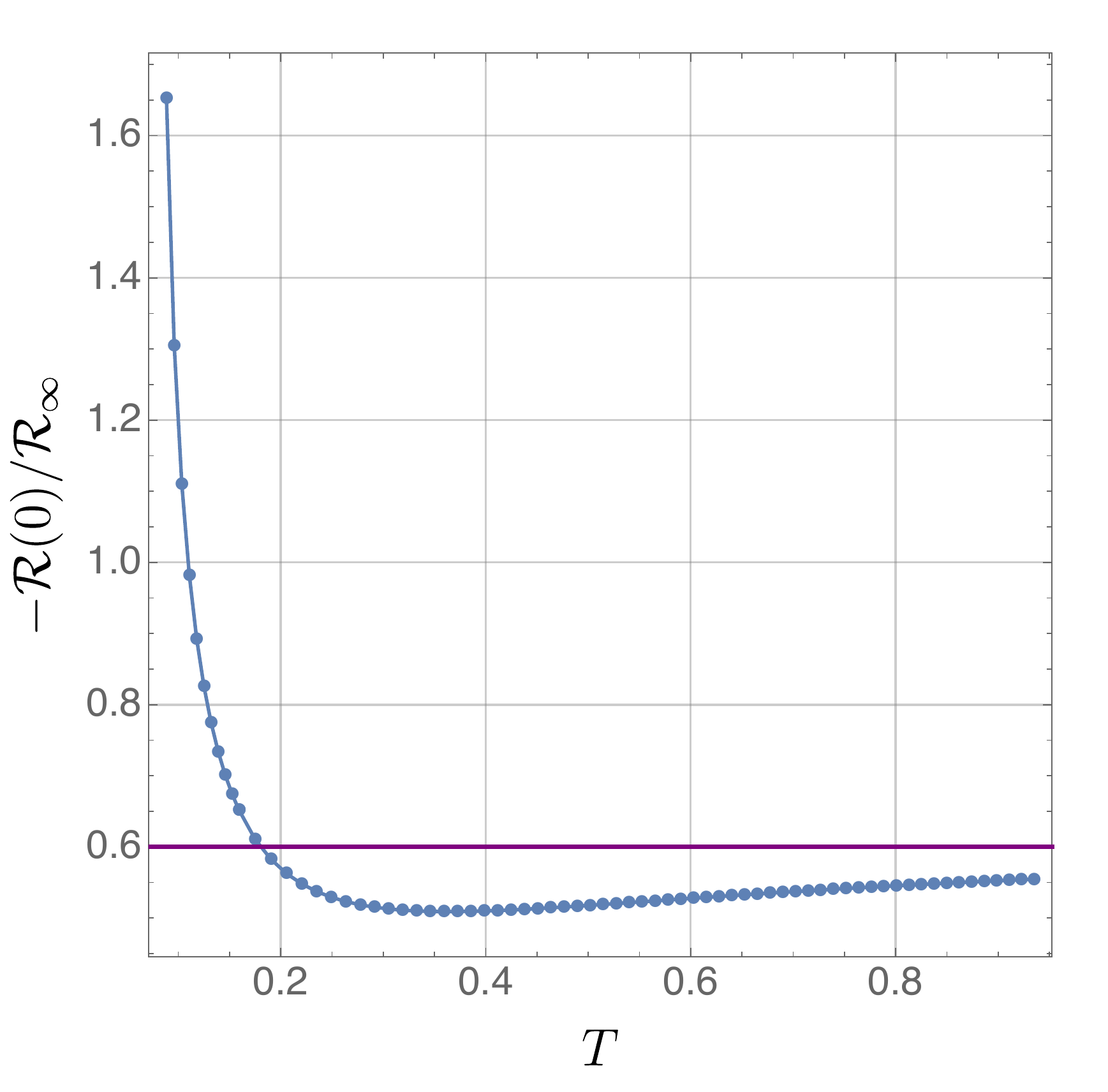}
 \caption{\label{fig:bulge2}Induced horizon curvature measured at the axis of rotation as a function of temperature computed for $\mu=4$ and $\nu=3$, with the horizontal purple line indicating the fluid/gravity prediction.}
 \end{center}
\end{figure}

In Fig.~(\ref{fig:bulge2}) we show the induced horizon curvature measured at the axis of rotation as a function of temperature (using $\mu=4$ and $\nu=3$), where we see the horizon becoming ever more curved as the temperature decreases. We believe that the extremal solution is singular, specially \emph{in lieu} of the results reviewed in \cite{Kunduri:2013ana} which rule out the existence of smooth extremal solutions other than the extremal uniform hyperbolic black holes. The horizontal purple line indicates the fluid/gravity result, which our novel solutions seem to approach at high temperature.

\begin{figure}[H]
\begin{center}
\includegraphics[width=.6\textwidth]{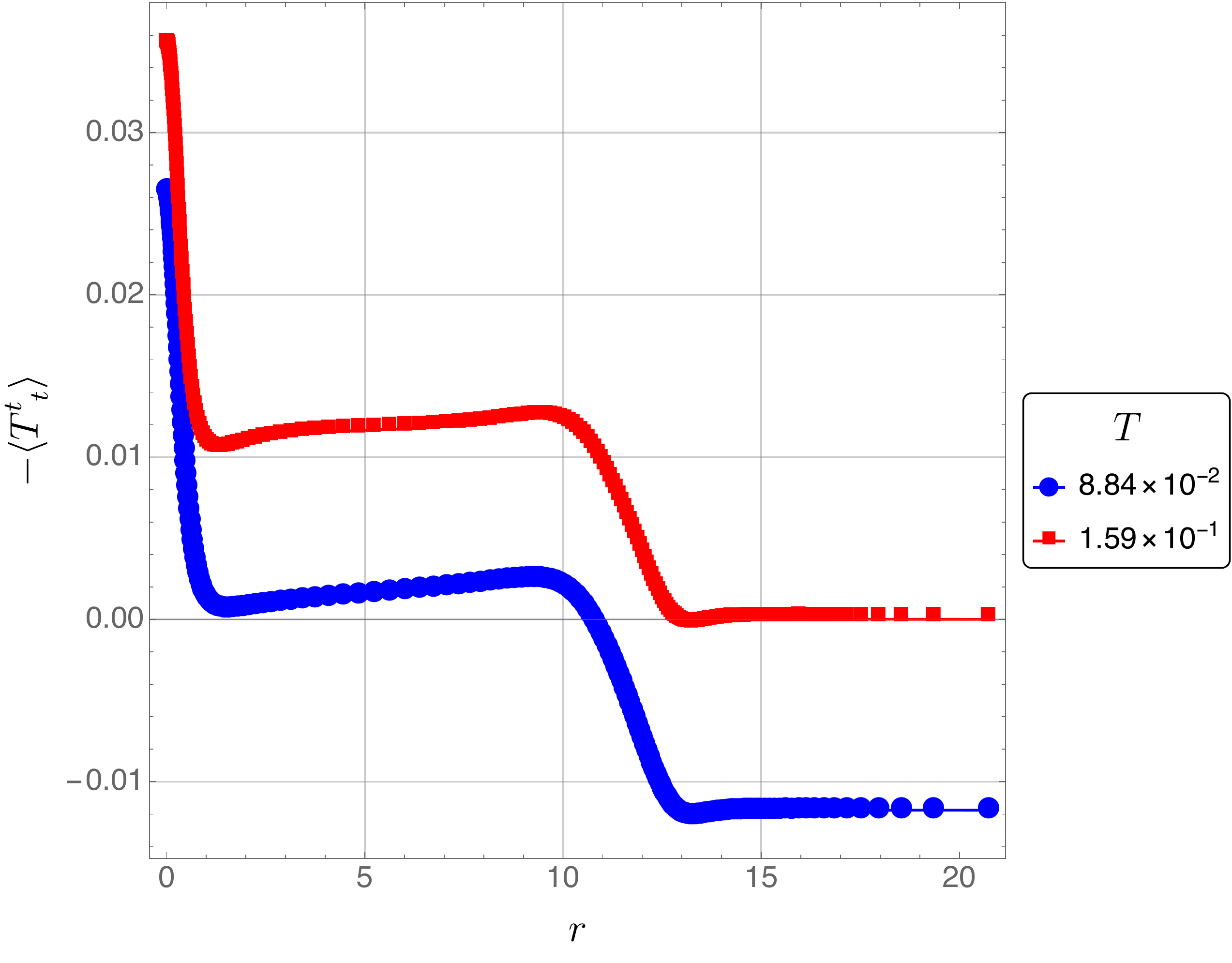}
 \caption{\label{fig:energy_density_wiggly}Energy density as a function of the boundary proper distance $r$ for the new black hole solutions, computed with $\mu=4$, $\nu=3$ (same parameters as in Fig.~\ref{fig:bulge}). The temperatures used are indicated in the legend above. (The upper curve has $T=1/(2\pi$).)}
 \end{center}
\end{figure}

Finally, we discuss the energy density of the novel solutions as a function of the boundary proper distance $r$, for several fixed temperatures $T$. This is an interesting quantity, which shows some unexpected features. In Fig.~\ref{fig:energy_density_wiggly} we plot the energy density, as a function of $r$, computed for two different temperatures and with $\mu=4$ and $\nu=3$. One curve appears to be a vertical shift of the other, with the lower temperature case being lower. There are  four distinct regimes, corresponding to different radial dependences, which we do not yet understand.

\subsection{Region D3}
The main change in moving from region D4 to D3 is that $F(y)$ acquires two  real zeroes in addition to $y=0$,  in the range
\be
-1 <y_-<y_+<0\,.
\ee
This means that in the region of spacetime containing the $y=0$ horizon, there is now a second horizon  at $y = y_+$. Both horizons are asymptotically hyperbolic, since the induced metric on the $y = y_+$ horizon is
\be \label{metric_yhorizon}
\mathrm{d}s^2 = \frac{1}{(x-y_+)^2} \left( \frac{\mathrm{d}x^2}{G(x)} + G(x) \mathrm{d}\phi^2 \right)
\ee
and its asymptotic region corresponds to $x \to y_+$. But $G(y_+) = F(y_+) +1$ and $  F(y_+) = 0$. So $G(y_+) =1$,  and after shifting the $x$ coordinate by the constant $y_+$, one sees that \eqref{metric_yhorizon}  approaches the standard hyperbolic metric.

The interpretation of this second horizon depends on the conformal frame chosen at infinity. If we continue to write the boundary metric in the form \eqref{withfactor}, it will have two asymptotically hyperbolic regions, one as $x \to 0$ and the other as $x\to y_+$. So it describes a wormhole.
This second asymptotic region can be viewed as the result of the  stretching we saw  in region D4 that  becomes infinite as one approaches the D3 boundary.  However, it is convenient to rescale the boundary metric to keep $x=y_+$ at a finite distance. This is accomplished by writing:
   \be
\mathrm{d}s^2_\partial = \frac{(1+\nu) }{4} \left(-\frac{F(x)}{x} \mathrm{d}t^2 + \frac{\mathrm{d}x^2}{xF(x) G(x)} -\frac{G(x)}{x} \mathrm{d}\phi^2 \right)\,,
\ee
This metric has the same asymptotic behavior as $x\to 0$ that we had before: {we introduce a rescaled time $\tilde{t} = (1+\nu)t/2$ so that for $x\rightarrow 0$ the metric approaches 
\be \mathrm{d}s^2_\partial = -\mathrm{d}\tilde{t}^2 +\mathrm{d}r^2 +\frac{e^{2r}}{4} \mathrm{d}\phi^2\,, \ee}
but now $x=y_+$ corresponds to a black hole on the boundary.  These are instances of \textit{black droplets}, first discovered in \cite{Hubeny:2009ru,Hubeny:2009kz}.

\begin{figure}[H]
\centering
{\includegraphics[width=.7\textwidth]{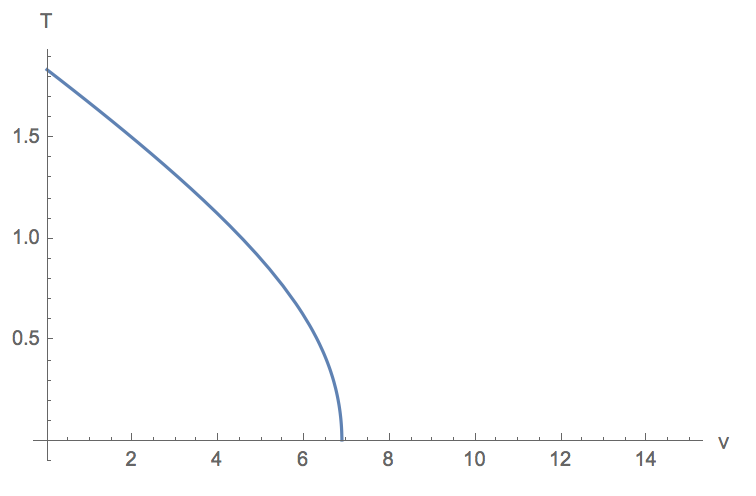} 
\caption{Temperature of the black droplet as a function of $\nu$ for fixed $\mu = 14.5$. The temperature goes to zero for $\nu = \mu -2\sqrt{\mu} \approx 6.88$.   \label{Fig10}}}
\end{figure}

 The metric above can be brought to the more usual form by the redefinition
\be
\frac{\mathrm{d}x}{2x\sqrt{G(x)}} = \mathrm{d}r
\ee
In this way we have
\be
\mathrm{d}s^2 = -\tilde{F}(x(r)) \mathrm{d}\tilde{t}^2 + \frac{\mathrm{d}r^2}{\tilde{F}(x(r))} -  \tilde{G}(x(r)) \mathrm{d}\phi^2
\ee
where we defined $\tilde{F}(x(r)) = \frac{1}{x(1+\nu)}  F (x(r)) $ and $\tilde{G}(x(r)) = \frac{(1+\nu)}{4x}  G (x(r)) $.
The temperature $T 
$ is given by
\be
T =  \frac{1}{4\pi} \frac{d \tilde{F}}{dr}\bigg{|}_{x=y_+} = \frac{1}{4\pi}  \frac{d \tilde{F}}{dx} \frac{dx}{dr}\bigg{|}_{x=y_+} 
\ee
A plot of this temperature is shown in Fig. \ref{Fig10}. We see from this plot that the temperature goes to zero as $\nu \rightarrow \mu -2 \sqrt{\mu}$, i.e., the boundary with D4. 

\begin{figure}[H]
{ \includegraphics[width=\textwidth]{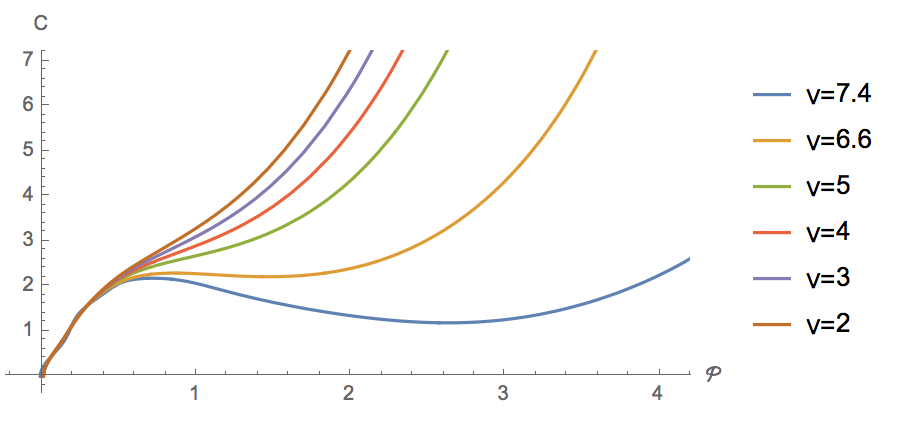}  \caption{Horizon geometry as function of the proper distance $\cal{P}$ for fixed $\mu = 14.5$. Increasing the parameter $\nu$ leads to  a globule with a longer neck on the horizon in the bulk. \label{Fig11}}}
\end{figure}

The black hole on the boundary can also induce  a globule on the horizon. Near extremality, a droplet with smaller temperature creates a globule with a longer neck. These results are shown in Fig \ref{Fig11} for configurations with fixed $\mu$, varying $\nu$.  

 The fact that the solutions in region D3 describe black droplets provides an interpretation for what the new solutions in D4 with growing horizon area might represent: they might approach a black funnel. In other words, the horizon should grow a giant bubble that eventually reaches the boundary. These might extend into region D3 as a  new class of black funnels which have a single connected horizon.  These new black funnel solutions remain to be constructed.
 
\subsection{\label{sec:hovering}Hovering black holes}

We now briefly discuss the possibility of hovering black holes. These are static, spherical black holes that hover above a noncompact horizon. (Charged versions were previously found in \cite{Horowitz:2014gva}.) A simple way to check whether hovering black holes are  possible is to look for a static timelike geodesic. If it exists, one expects that there are solutions with a small Schwarzschild black hole at the location of the geodesic.  Static timelike geodesics exist at any extremum of $g_{tt}$. By symmetry, it is natural to look along the axis of symmetry. For the solutions given by the C-metric, we have checked and indeed find static geodesics whenever $\nu\leq 1 - 3 \mu^{1/3} + \mu$.

In region D4,  these static geodesics are stable to small perturbations. They exist in an area close to the border with D3. There is an intuitive argument why one might expect hovering black holes to exist in this region.\footnote{We thank Don Marolf for suggesting this.} We have seen that near this border, the boundary geometry develops a large bulge near the axis. Since the horizon remains essentially unchanged, one can picture this part of the solution as analogous to global AdS. Since one can certainly put  a static spherical black hole in the center of global AdS, one might expect that inside this bubble region, a static spherical black hole should exist. This argument also suggests that if one deforms the spherical boundary of global AdS such that the equator shrinks to a narrow neck, there should be solutions containing off-center static spherical black holes. This is indeed the case \cite{Don}.

For the new solutions in region D4 containing horizons that grow with the boundary area, we do not find any static geodesics. This is consistent with the above picture, since part of the horizon already looks like an approximately spherical black hole inside the bulge on the boundary.
  
  The above condition for hovering black holes extends into region D3. So one could add a small static black hole between the asymptotically hyperbolic horizon and the black droplet. However these black holes  would be unstable, and would fall toward one of these two horizons if it is perturbed.

\subsection{\label{sec:D2Cmetric}Region D2}

The main difference in passing from D4 to D2 is that $G(x)$ acquires two real roots at $x_\pm$. Starting with the $x=-1$ axis, the solutions now have a second axis at $x = x_-$. This means that  the horizon  geometry \eqref{metric_horizon} becomes spherical. However, since the periodicity of $\phi$ is fixed by requiring that the $x=-1$ axis is smooth, there will usually be a conical singularity along $x=x_-$ with deficit angle:
\be
\delta \phi = \left|\frac{4\pi}{|G'(-1)|} - \frac{4\pi}{|G'(x_-)|}\right| = \frac{2 \pi }{1+\mu -\nu } \left(\frac{2 \mu -\nu }{\sqrt{\nu ^2-4 \mu }}-1\right)\label{consing} 
\ee
From \eqref{metric_horizon}, the area of this black hole is simply 
\be\label{bharea}
A = -(1 + 1/x_-) \Delta \phi = - \frac{4\pi ( 1 +x_-)}{x_-(1+\mu-\nu)}
\ee
Even though these black holes are accelerating, there is no acceleration horizon. This is possible in AdS for slow accelerations, and can be interpreted as spherical black holes which are held in place away from the origin. 

Since there is only one horizon at $y=0$, it is natural to take the boundary metric to be \eqref{eq:bdy0} which is the product of time and a curved two surface.  This two surface  will also be topologically a sphere with a conical deficit at $x = x_-$. Since we do not have an asymptotic region to compare to, there is no need to rescale by the extra constant factor as we did in \eqref{withfactor}. So the Hawking temperature is
$T = (1+\nu)/{4\pi}$ as in \eqref{hawkT0}.

These black holes were discussed in \cite{Chen:2015zoa}, where it was shown that as one approaches the border with D4, these black holes grow a long spike which eventually reaches infinity. However the area remains finite. It is easy to see that the same long spikes will appear in the boundary metric. This is because in both cases they arise because $G(x)$ is close to having a double zero.

 \begin{figure}[H]
  \begin{center}
{ \includegraphics[width=0.50\textwidth]{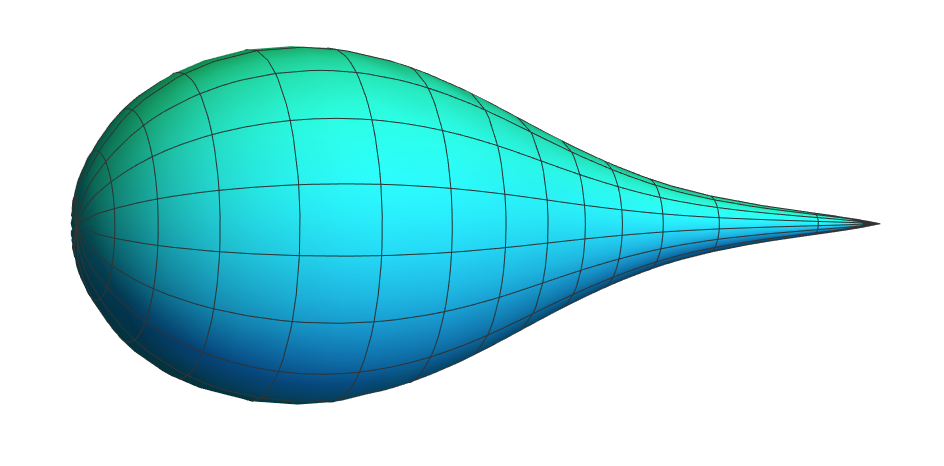} \includegraphics[width=0.29\textwidth]{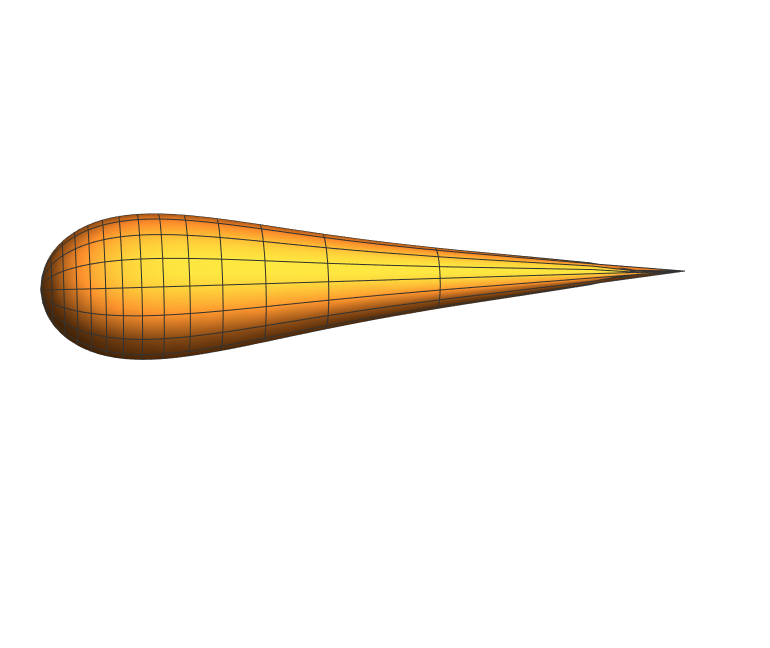}
 \caption{ Embedding in  Euclidean space of the boundary (left) and the horizon (right) of a ``bottle shaped" black hole of region D2 $\mu =13$, $\nu=7.23$, close to the border with D4: $\nu = 2\sqrt{\mu}$. \label{Fig:spike}}}
 \end{center}
\end{figure} 

However as we approach the border with D1, we again encounter a situation where the area of the boundary geometry grows without bound while the black hole remains essentially unchanged. This is because $F(y)$ again develops a double zero, and from \eqref{ord} it must lie between the two axes $-1< y_\pm<x_-$. In particular, the black hole area \eqref{bharea} has a finite limit at the D2/D1 interface since it is independent of $F$, while the area of the boundary geometry
\be
A= \Delta \phi \int_{-1}^{x_-} \frac{dx}{[-F(x)]^{3/2}}
\ee
clearly diverges there.
  Unlike the case in D4, the boundary is compact and its area is finite everywhere in D2. So the rapid change is area is easy to see  and does not need any regularization. Also the conical singularity \eqref{consing} changes smoothly  during this transition. Although these ``slowly accelerating" black holes in AdS have been well studied \cite{Krtous:2005ej,Appels:2016uha,Anabalon:2018ydc}, this feature has apparently not been noticed since it is not present in the usual parameterization in terms of $(m,A)$. {The new form of AdS C-metric parameterized in terms of $(\mu,\nu)$ provided in \cite{Chen:2015vma,Chen:2015zoa}} is more general.

This  would again pose a problem for gauge/gravity duality, if the black hole was described by a thermal state. We shall see in the next section that the spherical black holes described by the C-metric are analogous to ``small" AdS black holes, and  a new class of solutions exist which are analogous to ``large" black holes. In these new solutions, the boundary metric and bulk horizon behave similarly near the D2/D1 boundary.
\subsection{New solutions in D2}
In this section we do two things: 1) generalise the solutions of section \ref{sec:D2Cmetric} in such a way that the temperature can vary independently of the boundary metric; 2) find novel solutions with diverging horizon  area as the D2/D1 boundary is approached. While we could accomplish this using the same reference metric, it turns out that it is best (numerically) to do 1) and 2) using \emph{different} reference metrics.

Let us start with the proposed generalisation of section \ref{sec:D2Cmetric}. The idea is to use a reference metric that closely resembles the analytic solution described in \ref{sec:D2Cmetric}. However, the integration domain of such solutions is trapezoidal. While one can deal with such integration domains (see for instance \cite{Dias:2015nua}), here we perform a simple coordinate transformation that maps such trapezoid onto the unit square. We take the following coordinate transformation
\begin{equation}
x=(1+x_-)X-1\,\qquad y = [(1+x_-)X-1]\,Y^2\,,
\end{equation}
with $(X,Y)\in[0,1]^2$, the boundary is located at $Y=1$, the horizon at $Y=0$ and the poles at $X=0$ and $X=1$. In these coordinates, the metric becomes non-diagonal, but the integration domain is simple:
\begin{subequations}
\begin{multline}
\mathrm{d}s^2= \frac{1}{(1-Y^2)^2}\Bigg\{-Y^2\,q_1^{(0)}(X,Y)\mathrm{d}t^2+4 q_2^{(0)}(X,Y)\left[\mathrm{d}Y+Y\,q_3^{(0)}(X,Y)\mathrm{d}X\right]^2\\
+\frac{q_4^{(0)}(X,Y)\mathrm{d}X^2}{X(1-X)}+q_5^{(0)}(X,Y)\,X\,(1-X)\mathrm{d}\phi^2\Bigg\}\,
\end{multline}
where
\begin{align}
&q_1^{(0)}(X,Y)=\frac{x_-+x_+-x_- x_++\left(1-X-X x_-\right) \left(1-x_--x_+\right) Y^2-\left(1-X-X x_-\right){}^2 Y^4}{x_-\,x_+ \left[\left(1+x_-\right) X-1\right]}\,,
\\
&q_2^{(0)}(X,Y)=\frac{1}{A^{(0)}(X,Y)\left(1-X-X x_-\right)^2}\,,
\\
&q_3^{(0)}(X,Y)=\frac{1+x_-}{2 \left[\left(1+x_-\right) X-1\right]}\,,
\\
&q_4^{(0)}(X,Y)=\frac{x_- x_+}{\left(1-X-X x_-+x_+\right) \left[\left(1+x_-\right) X-1\right]^2}\,,
\\
&q_5^{(0)}(X,Y)=\frac{\left(1+x_-\right)^2 \left(1-X-X x_-+x_+\right)}{x_-\,x_+ \left[\left(1+x_-\right) X-1\right]^2}\,.
\end{align}
\label{eqs:cmetriccrazy}
\end{subequations}

The solution above can now be readily generalised to situations where the boundary metric is fixed, but where the temperature can vary independently. We take the following \emph{Ansatz} for the Einstein-DeTurck equation
\begin{multline}
\mathrm{d}s^2= \frac{1}{(1-Y^2)^2}\Bigg\{-Y^2\,q_1(X,Y)\mathrm{d}t^2+4 q_2(X,Y)\left[\mathrm{d}Y+Y\,q_3(X,Y)\mathrm{d}X\right]^2\\
+\frac{q_4(X,Y)\mathrm{d}X^2}{X(1-X)}+q_5(X,Y)\,X\,(1-X)\mathrm{d}\phi^2\Bigg\}\,
\end{multline}
with the reference metric being given by $q_1(X,Y)=q_1^{(0)}(X,Y)S(Y)$, $q_2(X,Y)=q_2^{(0)}(X,Y)/S(Y)$, $q_i(X,Y)=q_i^{(0)}(X,Y)$ for $i=3,4,5$ and $S(Y)=1+(1-Y^2)\alpha$.

We now briefly discuss the relevant boundary conditions. At the boundary, located at $Y=1$, we demand $q_i(X,1)=q_i^{(0)}(X,1)$ for $i\in{1,2,3,4,5}$. At the horizon, located at $Y=0$, regularity demands
\begin{equation}
q_1(X,1)q_2^{(0)}(X,1)=q_2(X,1)q_1^{(0)}(X,1)S(0)^2
\end{equation}
and $\left.\partial_Y q_i\right|_{Y=0}=0$ for $i\in\{1,2,3,4,5\}$, with the former Dirichlet condition implying the following Hawking temperature for our new solutions
\begin{equation}
T=\frac{(1+\nu) \left|1+\alpha\right| }{4 \pi }\,.
\end{equation}
Thus, for fixed boundary metric, \emph{i.e.} fixed $\mu$ and $\nu$, we can vary the temperature by dialing $\alpha$. The temperature corresponding to the solutions detailed in section \ref{sec:D2Cmetric} corresponds to setting $\alpha=0$. At $X=0$ we demand regularity which gives a Dirichlet condition relating $q_4(0,Y)$ and $q_5(0,Y)$
\begin{equation}
q_4(0,Y)q_5^{(0)}(0,Y)=q_5(0,Y)q_4^{(0)}(0,Y)
\end{equation}
and Robin boundary conditions for the remaining variables. Similar boundary conditions apply for $X=1$.

\begin{figure}[H]
  \begin{center}
 \includegraphics[width=0.50\textwidth]{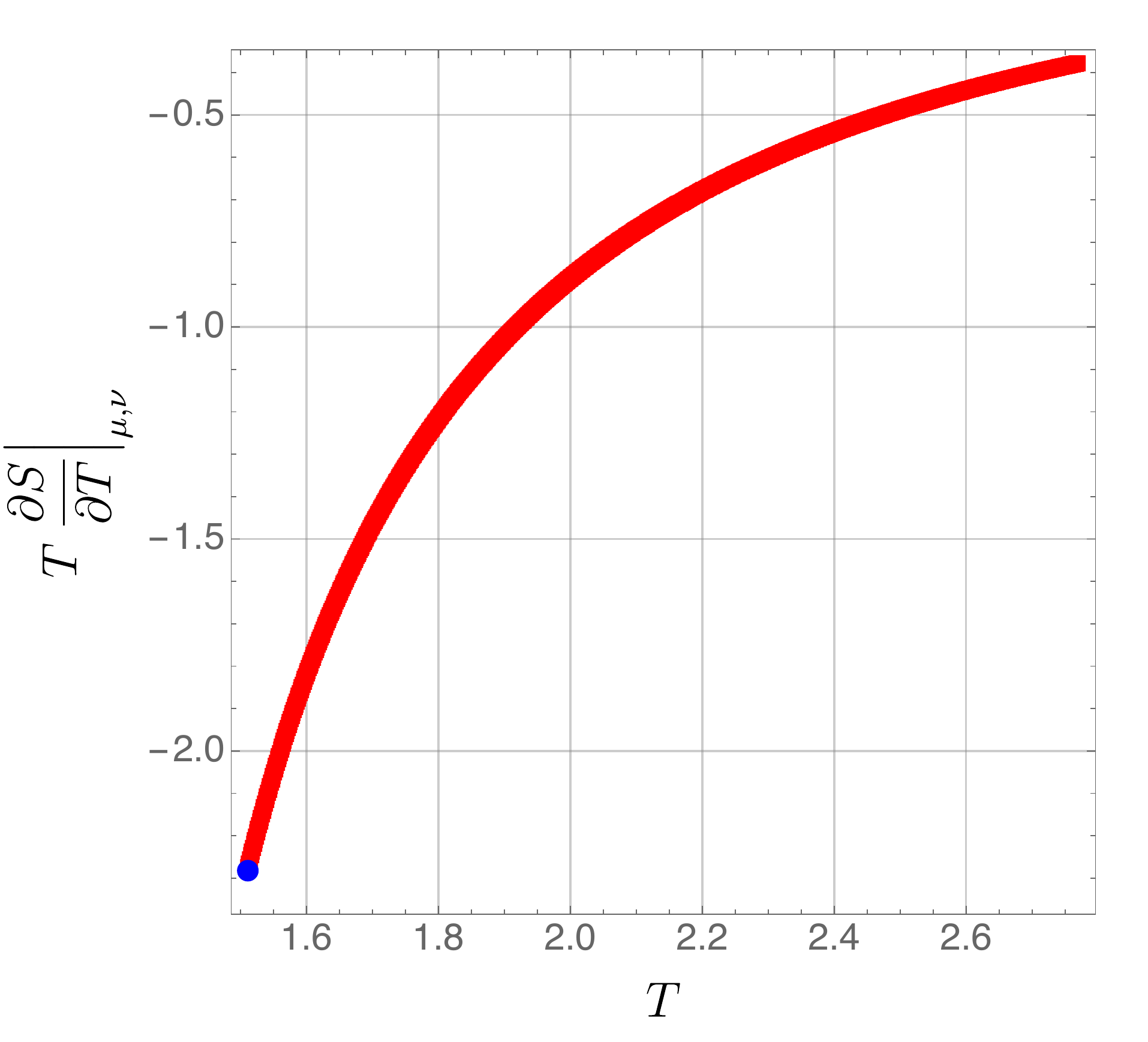}
 \caption{\label{fig:specificheat}Specific heat of the C-metric like solution computed for fixed boundary metric with $\mu=\nu=18$. The blue disk denotes the C-metric solutions of section \ref{sec:D2Cmetric}, while the red squares represents generalisations to different temperature.}
 \end{center}
\end{figure}

This new \emph{Ansatz} allow us to compute the specific heat of the solutions described in section \ref{sec:D2Cmetric}. We find that the specific heat we compute at $T=(1+\nu)/(4\pi)$ is always negative for the C-metric like solutions, and thus they cannot describe a thermal state in the dual theory. In Fig.~\ref{fig:specificheat} we plot the specific heat for $\mu=\nu=18$ as a function of $T$, with the blue disk indicating the solutions described in section \ref{sec:D2Cmetric}, and the red squares its generalisation to arbitrary temperatures.

Next, we constructed novel solutions with the same boundary metric as those in section \ref{sec:D2Cmetric} and, in particular, which behave very differently near the D2/D1 boundary. These solutions never seem to merge as solutions, that is to say as a second order phase transition, with the C-metric solutions we found. Although we found some of these new solutions using the metric \emph{Ansatz} corresponding to Eqs.~(\ref{eqs:cmetriccrazy}), our numerical methods exhibited very poor convergence. In order to solve this, we changed our reference metric. We take the following metric \emph{Ansatz}
\begin{multline}
\mathrm{d}s^2=\frac{1}{\left(1-\tilde{y}^2\right)^2} \Bigg\{-\tilde{y}^2 G_{\tilde{y}}(\tilde{y}) q_1(w,\tilde{y}) \mathrm{d}t^2+\frac{4 y_+^2 q_2(w,\tilde{y})}{G_{\tilde{y}}(\tilde{y})}\left[\mathrm{d}\tilde{y}+q_3(w,\tilde{y})\mathrm{d}w\right]^2+\\
   \frac{y_+^2}{F_w(w)} \left[\frac{4 q_4(w,\tilde{y}) \mathrm{d}w^2}{2-w^2}+G_w(w) (1-w^2)^2 q_5(w,\tilde{y}) \mathrm{d}\phi^2\right]\Bigg\}\,,
   \label{eq:giant1}
\end{multline}
with
\begin{subequations}
\begin{align}
&G_{\tilde{y}}(\tilde{y})=\left(1-\tilde{y}^2\right)^2+\left(3-3 \tilde{y}^2+\tilde{y}^4\right) y_+^2\,,
\\
&G_w(w)=\frac{1}{4} \left[x_+-T(w)\right]^2 T(w)\,\tilde{F}(T(w)) \mu ^2 \left(1+x_-\right)^2\,,
\\
&F_w(w)=\mu  T(w)^2 \left[x_+-T(w)\right]\,\tilde{F}(T(w))^2\,,
\\
&T(w)=\frac{1}{2} \left(1+x_-\right) w \sqrt{2-w^2}-\frac{1}{2} \left(1-x_-\right)\,,
\\
&\tilde{F}(X)=-\left[1+\nu +(\mu +\nu )\,X+\mu\, X^2\right]\,.
\end{align}
\end{subequations}
For the reference metric we choose $q_1=q_2=q_4=q_5=q_3+1=1$. Although this \emph{Ansatz} seems more complicated, it captures more easily the solutions we wish to find. We now briefly discuss boundary conditions. The conformal boundary is located at $\tilde{y}=1$, where we demand
\begin{equation}
q_1=q_2=q_4=q_5=q_3+1=1\,.
\end{equation}
Note that $G_{\tilde{y}}(1)=y_+^2$, which brings the metric to the following form
\begin{multline}
\left.\mathrm{d}s^2\right|_{\tilde{y}\simeq 1}=\frac{1}{4\left(1-\tilde{y}\right)^2} \Bigg\{-y_+^2\, \mathrm{d}t^2+4\mathrm{d}\tilde{y}^2+\frac{y_+^2}{F_w(w)} \left[\frac{\mathrm{d}w^2}{2-w^2}+G_w(w) (1-w^2)^2 \mathrm{d}\phi^2\right]\Bigg\}\,.
\end{multline}
Taking now
\begin{equation}
\tilde{y}=1-\frac{y_+}{2}\,z\quad\text{and}\quad x=\frac{1}{2} \left(1+x_-\right) w \sqrt{2-w^2}-\frac{1}{2} \left(1-x_-\right)\,,
\end{equation}
yields
\begin{equation}
\left.\mathrm{d}s^2\right|_{z\simeq 0}=\frac{1}{z^2} \Bigg\{-\mathrm{d}t^2-\frac{1}{F(x)} \left[-\frac{\mathrm{d}x^2}{F(x)G(x)}+G(x) \mathrm{d}\phi^2\right]+\mathrm{d}z^2\Bigg\}\,,
\end{equation}
which is exactly the desired boundary metric. At the horizon, located at $\tilde{y}=0$, regularity, together with our choice of reference metric, implies
\begin{equation}
q_1(w,0)=q_2(w,0)\,,
\end{equation}
which in turn gives for the Hawking temperature of our new solutions
\begin{equation}
T = \frac{1+3\,y_+^2}{4\pi\,y_+}\,.
\end{equation}
This shows that $y_+$ controls the background temperature of our solutions. If we wish to find solutions exactly with the same temperature and boundary metric as the solutions we found in section \ref{sec:D2Cmetric} we further need to set
\begin{equation}
T=\frac{1+3\,y_+^2}{4\pi\,y_+}=\frac{1+\nu}{4\pi}\Rightarrow y_+ = \frac{1}{6} \left(1+\nu +\sqrt{\nu ^2+2 \nu -11}\right)\,.
\label{eq:ticktack}
\end{equation}
For the remaining variables we find $\partial_{\tilde{y}}q_1=\partial_{\tilde{y}}q_2=\partial_{\tilde{y}}q_4=\partial_{\tilde{y}}q_5=q_3=0$ at $\tilde{y}=0$.

The line element (\ref{eq:giant1}) further has two singular ``points'' where we need to provide boundary conditions. These correspond to points where the circle parametrised by $\phi$ shrinks to zero, \emph{i.e.} $w=\pm1$, which we will loosely call poles. Here, our choice of reference metric and regularity gives $q_5(\pm1,\tilde{y})=q_4(\pm1,\tilde{y})$ and $\partial_w q_1=\partial_w q_2=\partial_w q_4=\partial_w q_5=q_3=0$ at $w=\pm1$.

Our new \emph{Ansatz} was also able to reproduce the C-metric solutions of section \ref{sec:D2Cmetric}, but needed a much higher number of points to achieve the desired accuracy. On the other hand, we were able to find new solutions relatively easily. These  new solutions behave in a markedly different manner to the C-metric solutions as we approach the D2/D1 interface as we now show.

\begin{figure}[H]
  \begin{center}
 \includegraphics[width=0.50\textwidth]{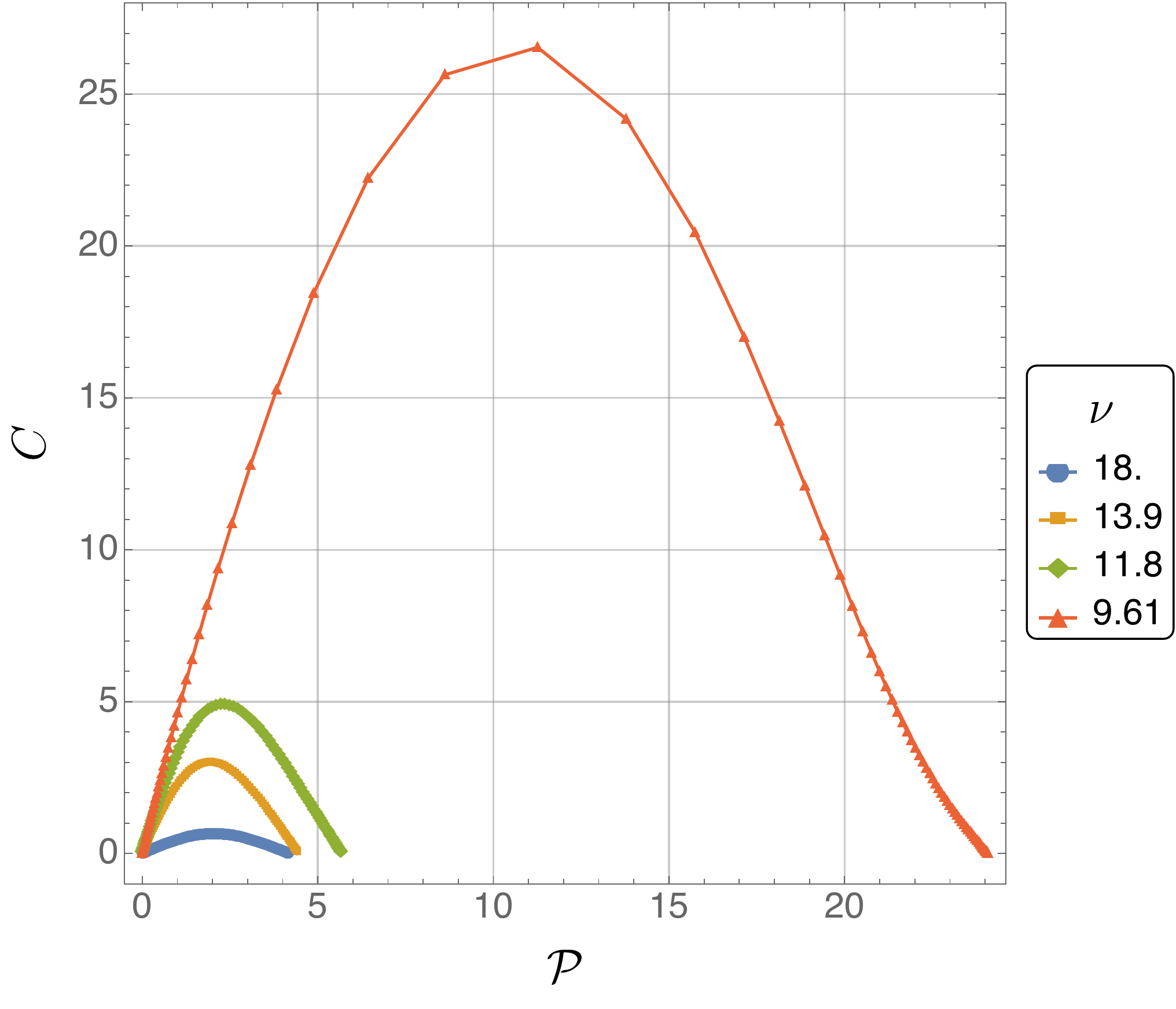}
 \caption{\label{fig:greatphi}The circumference $C$ of the $\phi$ circles  as a function of  proper distance $\mathcal{P}$ for fixed $\mu=18$ and using $y_+$ as in Eq.~(\ref{eq:ticktack}) for the new solutions: different symbols correspond to difference choices of $\nu$. For $\mu=18$, the D2/D1 interface is located around $\nu\approx 9.51472$.}
 \end{center}
\end{figure}

\begin{figure}[H]
  \begin{center}
 \includegraphics[width=0.48\textwidth]{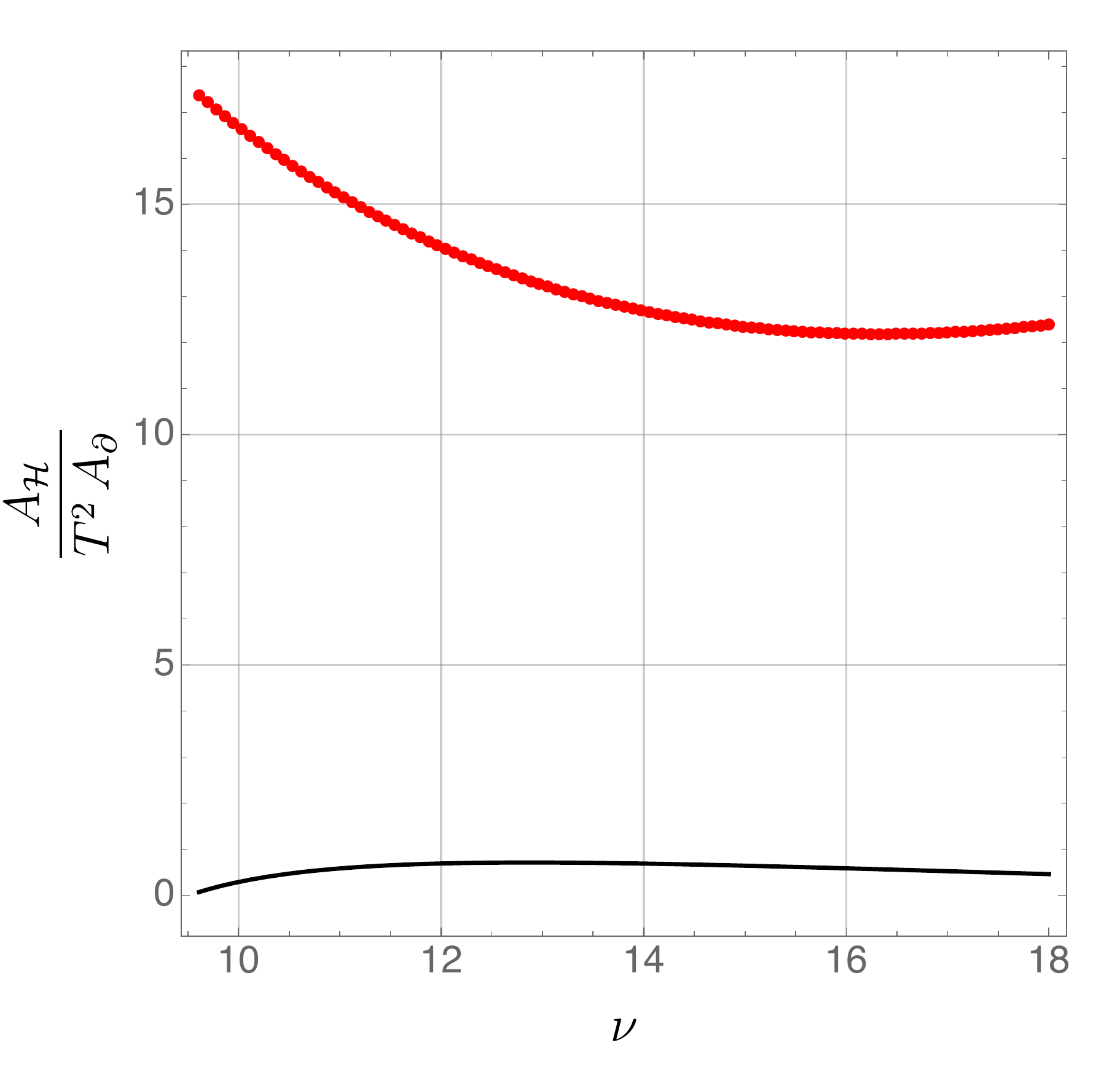}
 \caption{\label{fig:ratioglobal}The ratio $A_{\mathcal{H}}/(T^2\,A_{\partial})$ as a function of $\nu$ for fixed $\mu=18$ and using $y_+$ as in Eq.~(\ref{eq:ticktack}) to fix the temperature: the red disks correspond to our new numerical solutions, and the black solid line shows the corresponding quantity computed for the C-metric solutions of section \ref{sec:D2Cmetric}. For $\mu=18$, the D2/D1 interface is located around $\nu\approx 9.51472$.}
 \end{center}
\end{figure}

 First, the horizon of the new solutions grows in size as we approach the D2/D1 boundary. This can be seen in  Fig.~\ref{fig:greatphi} where we plot the circumference $C$ of the  $\phi$ circles as a function of the proper distance $\mathcal{P}$ along the black hole horizon from the pole at $w=-1$. In this particular plot we take $y_+$ as in Eq.~(\ref{eq:ticktack}) and $\mu=18$, but different choices give qualitatively similar results, so long as our chosen value of $\mu$ allow us to approach the D2/D1 interface. Different symbols in Fig.~\ref{fig:greatphi} represent different choices for $\nu$, which are labelled in the corresponding figure. We see that as the boundary between D2/D1 is approached, the horizon grows rapidly.
  We next want to compare the growth of the horizon area $A_{\mathcal{H}}$ with the growth of the area of the boundary, $A_{\partial}$. Since the entropy on the boundary is proportional to $T^2\,A_{\partial}$ and the entropy in the bulk is proportional to $A_{\mathcal{H}}$, we compute their ratio and find that it  remains finite at the interface.  This can be seen in Fig.~\ref{fig:ratioglobal} where we keep $y_+$ given as in Eq.~(\ref{eq:ticktack}) to determine $T$, fix $\mu=18$ and vary $\nu$. The red disks correspond to our new numerical solutions, and the black solid line shows the corresponding quantity computed for the C-metric solutions of section \ref{sec:D2Cmetric}, which vanishes at the D2/D1 interface (for $\mu=18$, the D2/D1 interface is located at $\nu\approx 9.51472$).

\begin{figure}[H]
  \begin{center}
 \includegraphics[width=0.55\textwidth]{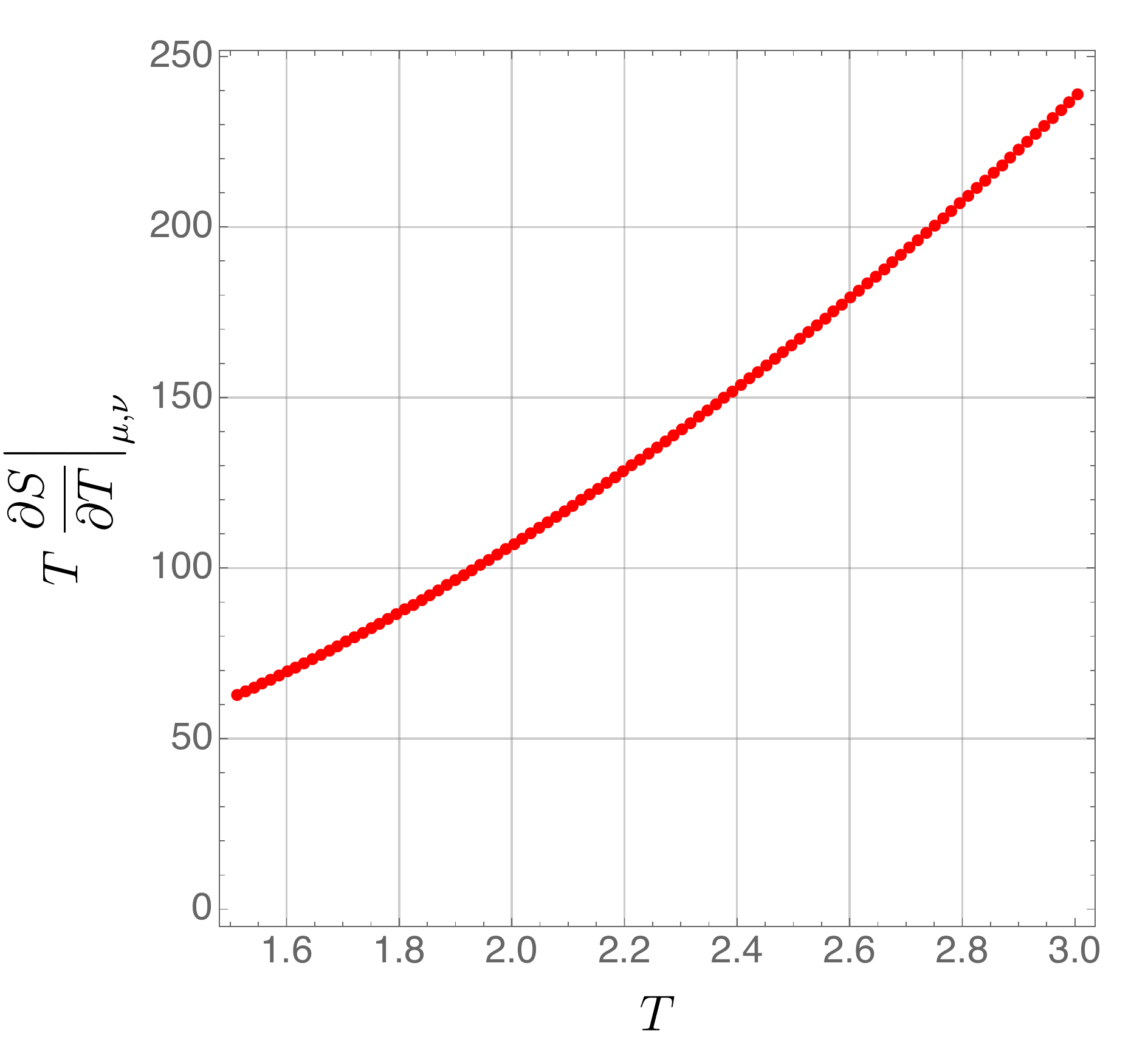}
 \caption{\label{fig:specificheatglobalnew}Specific heat of the new solutions as a function of their temperature, computed for fixed boundary metric with $\mu=\nu=18$.}
 \end{center}
\end{figure}

Finally, we have also computed the specific heat associated with our new solutions for fixed boundary parameters. In contrast to the C-metric solutions, it turns out to be positive, as seen in Fig.~\ref{fig:specificheatglobalnew} for $\mu=\nu=18$.

Although the new solutions we have constructed satisfy many of the requirements to be dual to a thermal state, they may not actually dominate the canonical ensemble. This is because they contain a conical singularity extending from the boundary into the bulk. It has been shown in many cases that there exist smooth bulk solutions that have boundary metrics with conical singularities. Furthermore the energy of the smooth solution is usually lower than the energy of the singular one. A simple example is  the ground state of a static cone: $-dt^2 + dr^2 + r^2 d\phi^2$ where the period of $\phi$ is less than $2\pi$.  One bulk solution is the Poincar\'e patch of AdS with this new period of $\phi$, which has   a conical singularity extending to the Poincar\'e horizon. However a nonsingular bulk solution with the same boundary (conformal) metric can be obtained by rescaling:
\be
-\mathrm{d}t^2 + \mathrm{d}r^2 + r^2 \mathrm{d}\phi^2 = r^2 [  \frac{-\mathrm{d}t^2 + \mathrm{d}r^2 }{r^2} + \mathrm{d}\phi^2]
\ee
So the boundary is conformally equivalent to AdS$_2 \times S^1$. One can now take a hyperbolic black hole and analytically continue $ t \rightarrow i\phi$ and $H_2 \rightarrow \mathrm{AdS}_2$ to get a smooth bulk solution with the same boundary geometry. The mass parameter in the solution  is fixed by the period of $\phi$.  (The resulting spacetime does not have a black hole.)  One finds that this solution has negative energy density  \cite{Hickling:2014dra}. Other examples can be found in \cite{Horowitz:2017ifu}. 

This suggests that the dominant bulk configuration at high temperature will be a {\it smooth} black hole with horizon area that grows with the area of the boundary geometry. Finding this new class of solutions remains an open problem.
 \section*{Acknowledgments} 
It is a pleasure to thank Xi Dong and Don Marolf for discussions. GH was supported in part by NSF grants PHY-1504541 and PHY-1801805. CT acknowledges support from the NSF Grant PHY-1125915 and the Saclay-Polytechnique ANR grant: Black-dS-String (ANR-16-CE31-0004). JES were supported in part by STFC Grants No. PHY-1504541 and ST/P000681/1.
\appendix
\section{Appendix: Generalization to nonzero electromagnetic charges}
The neutral AdS C-metric that we have analyzed in the previous sections can be easily generalized by the addition of electric and magnetic charges \cite{Plebanski:1976gy}. The metric for the charged case is of the form \eqref{C-metric}: the structure functions $F(y)$ and $G(x)$ are generalized 
to
\be \label{Fcharged}
F(y) = y [1+ \nu + (\mu +\nu)y + (\mu -z^2) y^2-z^2 y^3 ]
\ee
\be \label{Gcharged}
G(x) = (1+x) (1+ \nu x + \mu x^2 -z^2 x^3)
\ee
where $z^2 = e^2 + g^2$. The gauge field $A$ reads
\be
A = l (e y dt -g (1+x) d \phi)\,.
\ee  
Notice that $F,G$ are now polynomial of fourth order in $y$ and $x$ respectively. Moreover, once again the relation $G(x) -F(x) =1$ holds. For simplicity, we restrict to the case of purely electric charge: throughout this section we will set $g=0$. 

 Notice that the asymptotic form of both the horizon and boundary geometries are unaffected by the addition of these charges. As in the uncharged C-metric \eqref{C-metric}, there is a horizon at $y=0$ and the boundary is at  $x=y$. The asymptotic region of each corresponds to
 $x\rightarrow 0$, and in this limit $G=1$ and $F=x(1+\nu)$, so the presence of charges affects only  the geometry in the interior.
 
As before, there is an axis at $x=-1$. In order to avoid conical singularities on the axis the angular coordinate $\phi$ should be identified with periodicity
\be
\Delta \phi = \frac{4 \pi}{ |1+\mu-\nu +z^2|}
\ee
Moreover, the  temperature of the horizon at $y=0$ (with respect to $\partial_t$) is
\be
T= \frac{F'(0)}{4 \pi} = \frac{1 + \nu}{4 \pi}
\ee
which, interestingly, does not depend on the charge parameter $z$.

The domain analysis of this class of configurations was initiated in \cite{Chen:2015vma}, where it was shown that the possible domains are again box-like shaped, trapezoidal and triangular, in analogy with the uncharged cases (a)-(b)-(c)-(d) of Figure \ref{Fig9}.  We would like here to build upon their classification: in particular we are going to map the parameter space of the various regions in some examples, and analyze the boundary geometry and the behaviour of the chemical potential for deformed hyperbolic  black holes.

\begin{figure}[H]
\begin{center}
 {\includegraphics[width=.6\textwidth]{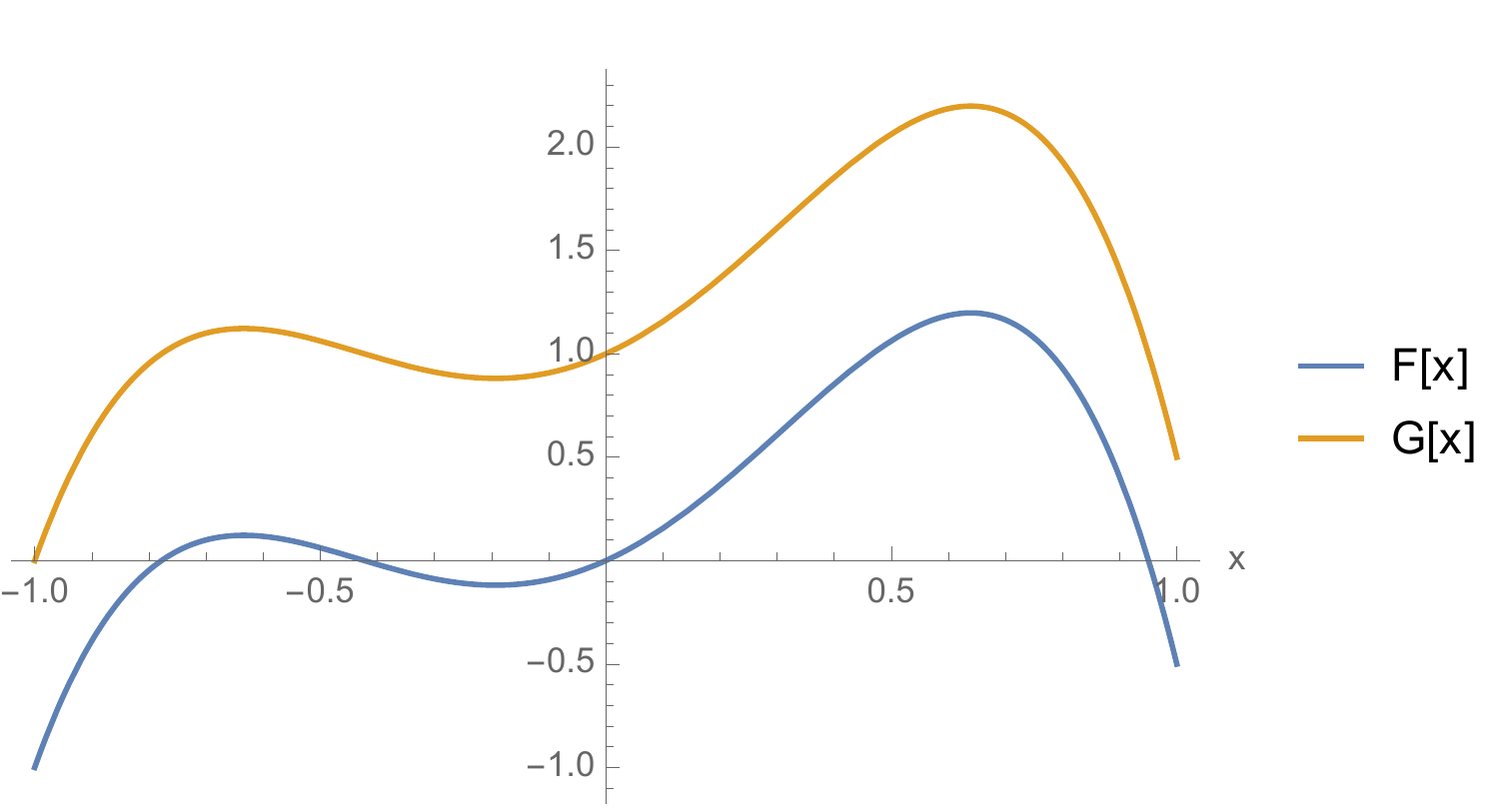}
\caption{\label{Fig23} Plot of $F(x)$ and $G(x)$ for the values $z=2,\mu=3,\nu=1/4$.}}
 \end{center}
\end{figure}

Let us start with the fact that $F(y)$ and $G(x)$ are polynomials of degree four, which can be cast in the following form
\be
f(u) = -z^2 (u-u_1)(u-u_2)(u-u_3)(u-u_4)\,.
\ee
From\eqref{Fcharged}-\eqref{Gcharged} one can see that $F(y)$ has a root at $y=0$ and $G(x)$ has one at $x=-1$. The other roots are given by solving two third-order polynomials and they are not particularly illuminating, therefore we do not report them here. It is immediate to see that the functions $F$ and $G $ go to $-\infty$ for $x \rightarrow \pm \infty$ and their possible schematic behavior is depicted in Fig. \ref{Fig23}.
From Fig. \ref{Fig23} one can see that the order of the roots, in the case where all of them are real, is
\be \label{order_charged}
x_1<y_1< y_2 < x_2 <x_3 <y_3 < y_4 < x_4 \,.
\ee
Our aim here is to generalize the domains of type D found in the uncharged case to the presence of additional charge. In similarity to the uncharged case, we take Region D characterized by the coordinate range $-1<x<y<0$. In particular $x=-1$ should be a left root and, in order to have Lorentzian signature, the function $G$ needs to be positive throughout  the interval, so the $G$ root at $x=-1$ needs to have positive derivative. The condition $G'(-1)>0$ is equivalent to
\be
1 + \mu - \nu + z^2>0\,,
\ee
and that is the region we will be interested in. In both cases either $x_1=-1$ or $x_3=-1$, the spacetime of interest cuts off at $x=-1$.
Now there are two cases: either $F$ is decreasing or increasing at $y=0$. In the first case, $F'(0) < 0$ implies  $ \nu<-1$ while in the second case 
$F'(0) > 0$ implies $ \nu>-1$.
Notice that this second condition is the one used for the parameter space of the uncharged solution. Only in this second case does the region D4 appear. In this case, either $y_3=0$ (and $y_1$ and $y_2$ are complex) or $y_1=0$. Since we are mostly interested in regions D1-D2-D3-D4, we restrict in what follows to this second case, hence our plots will have the following range of parameters\footnote{For $\nu<-1$ there are solutions which have qualitatively the same domain structure as region D3. Moreover, in this section, since we do not consider real roots of $F$ and $G$ which lie outside the region $-1<x<y<0$, we do not distinguish region B, C and region D4, and we collectively denote it with D4.}
\be
1 + \mu - \nu + z^2>0\,, \qquad \qquad
\nu>-1\,.
\ee
Notice that in this case the additional fourth roots that appear due to the presence of electric charge do not modify the causal structure, since they lie outside the range of interest.

\begin{figure}[H]
\begin{center}
 {\includegraphics[width=.48\textwidth]{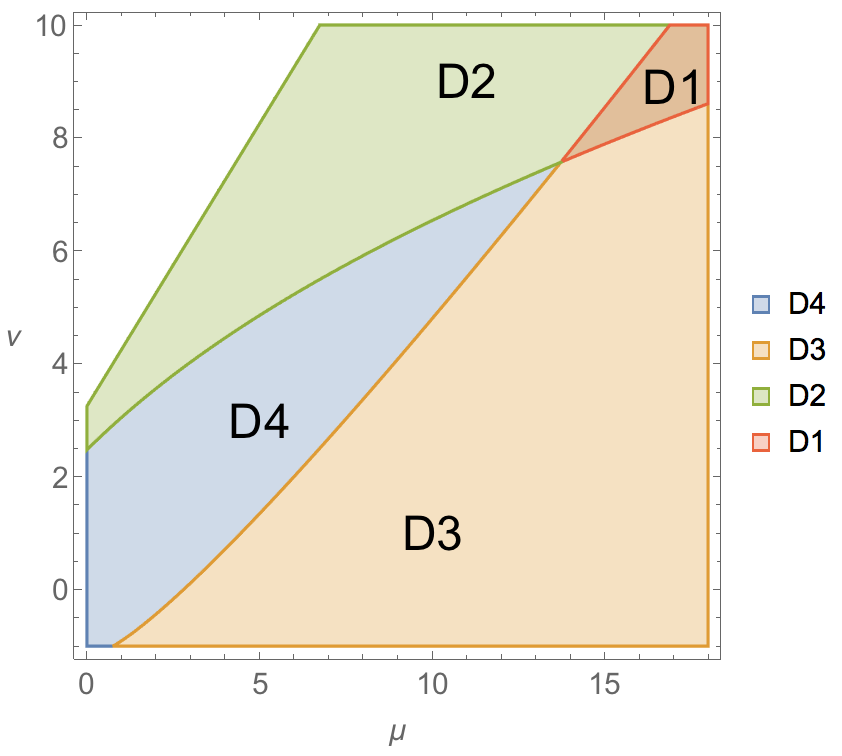}\includegraphics[width=.525\textwidth]{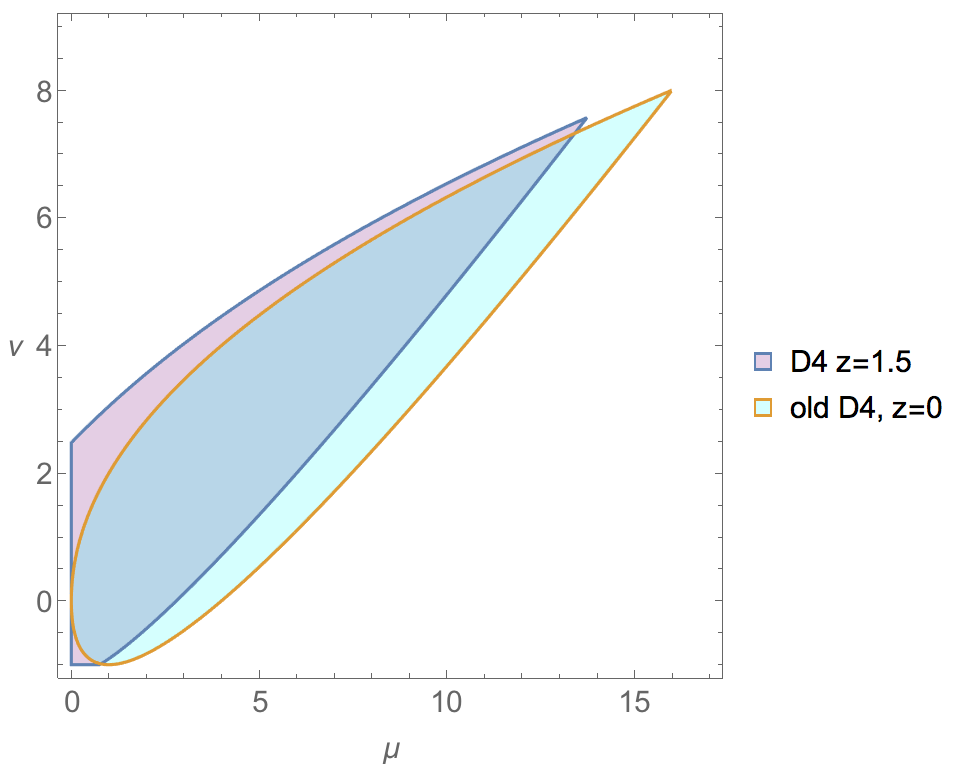}
\caption{\label{Fig24} On the left, plot depicting the various parameter space for regions D1-D4 for fixed $z=1.5$. On the right, a comparison of the size of region D4, for $z=0$ and $z=1.5$.}}
 \end{center}
\end{figure}

 It is useful at this point to recap from the previous sections the characterization of regions D1-D4. In addition to the known roots at $x=-1$ and $y=0$ (and ignoring the fourth root) the behavior of the remaining two roots is as follows:
\begin{itemize}
\item in region D1 there are two real roots of $F(y) $ and two of $G(x)$
\item  in region D2 there are two real roots of $G(x)$
\item  in region D3 there are two real roots of $F(y)$
\item  in region D4 there are no real roots
\end{itemize} 
This gives rather complicated conditions on the phase space of parameters. It is instructive to plot a section of the phase space diagram with fixed, nonzero charge $z$ (and $\mu>0$) to see how the regions D1, D2, D3, D4 change upon addition of electric charge. In particular, 
if we define  the following quantities:
\be
A= (2 \mu^3 + 9 \mu \nu z^2 + 27 z^4)^2 -4 (\mu^2 + 3 \nu z^2)^3 
\ee
and 
\be
B=(3 \mu z^4-2 \mu^3 - 3 \mu^2 z^2 - 9 \mu \nu z^2 - 27 z^4  - 18 \nu z^4 + 
   2 z^6)^2 - 4 (3 (\mu + \nu) z^2 + (\mu - z^2)^2)^3
\ee
we find that region D1 is bounded in the $\mu-\nu$ slice by the following expressions
\be
A<0 \qquad \text{and} \qquad B<0
\ee
D4 is bounded by \be
A>0 \qquad  \text{and} \qquad B>0
\ee
Moreover, region D2 is bounded above by the following curve
\be
1+\mu-\nu+z^2>0\,
\ee
and below by D4 and D1. Finally, region D3 is bounded above by D4 and D1 and below by $\nu=-1$. 
In Fig. \ref{Fig24},  the four regions are depicted in the parameter space $\mu, \nu$, keeping $z=1.5$ fixed. A comparison of the ``old" ($z=0$) D4 region and the $z=1.5$ region is also provided. One can see that the addition of electric charge shifts the D4 region to the left and makes it wider when going towards lower values of $\mu$.

In what follows we are interested in particular in deformed hyperbolic black holes, which correspond to region D4 in Fig. \ref{Fig24}. Similarly to the uncharged case, the warp functions $F(y)$ and $G(x)$ in that case do not have any additional real roots and the spacetime is characterized by the usual coordinate range $-1<x<y<0$. The metric on the $y=0$ horizon is again given by \eqref{metric_horizon} and the metric on the boundary is \eqref{withfactor}. In particular, we need to perform a constant rescaling of $t$ so that the temperature becomes $T = 1/(2\pi)$ as before.

\begin{figure}[H]
\begin{center}
 {\includegraphics[width=.068\textwidth]{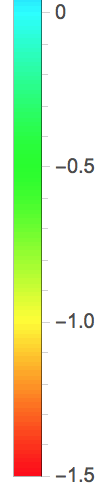}  \includegraphics[width=.32\textwidth]{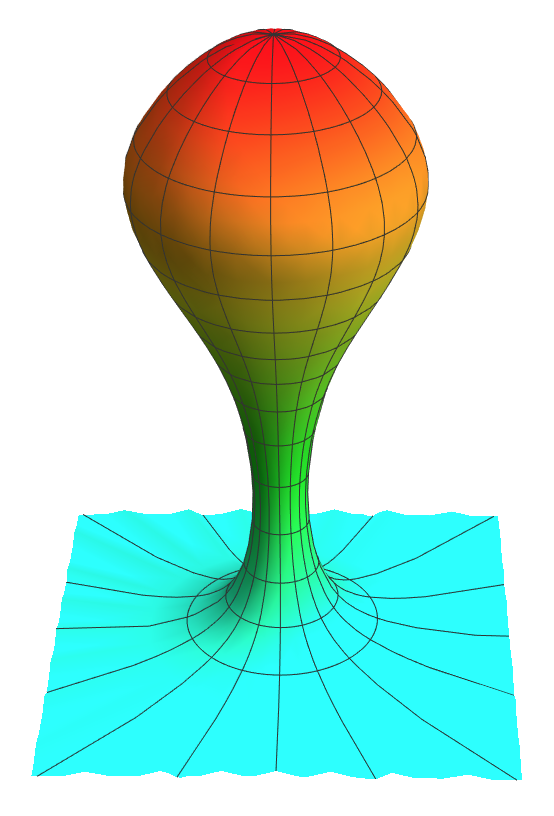}  
 \includegraphics[width=.2\textwidth]{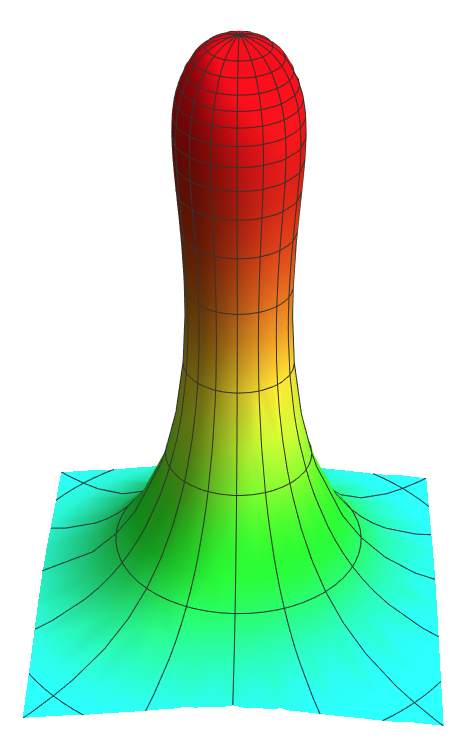} 
\caption{\label{Fig25} This is a globule on the boundary (left), and on the horizon (right) for the values $\mu=10,\ z=1.5, \ \nu=6$, embedded in hyperbolic space. The chemical potential $\Phi$ (on the left) and the electric field $\mathcal{E} =-\sqrt{|F_{yt}F^{yt}|}$ (right), whose maximum denoted by the color red in the figure, occur at the tip  (``spherical part") of the globule.  }}\end{center}
\end{figure}

These deformed hyperbolic black holes, as we might expect, may present black globules on their horizon. It is interesting to study such configurations in the presence of charge, because the chemical potential $\Phi$, computed as the difference between the value of the gauge field at the boundary and that at the horizon
\be
\Phi = A_t|_{y=x} - A_t|_{y=0} = A_t|_{y=x} = l e x
\ee
depends on the angular coordinate $x$ - it is in fact linear in $x$. It reaches a maximum value at $x=-1$ hence towards the center where the globule is present, and it decreases towards the asymptotic hyperbolic space. We therefore have an inhomogeneous chemical potential on the boundary, where a charged defect is present. The situation is similar to the setup studied in \cite{Horowitz:2016ezu}, where a spatially-dependent chemical potential induced mushroom-like structures on a planar horizon. We can plot the globule on the boundary, and the corresponding horizon geometry, obtaining Fig. \ref{Fig25}.

\begin{figure}[H]
\begin{center}
 {\includegraphics[width=.5\textwidth]{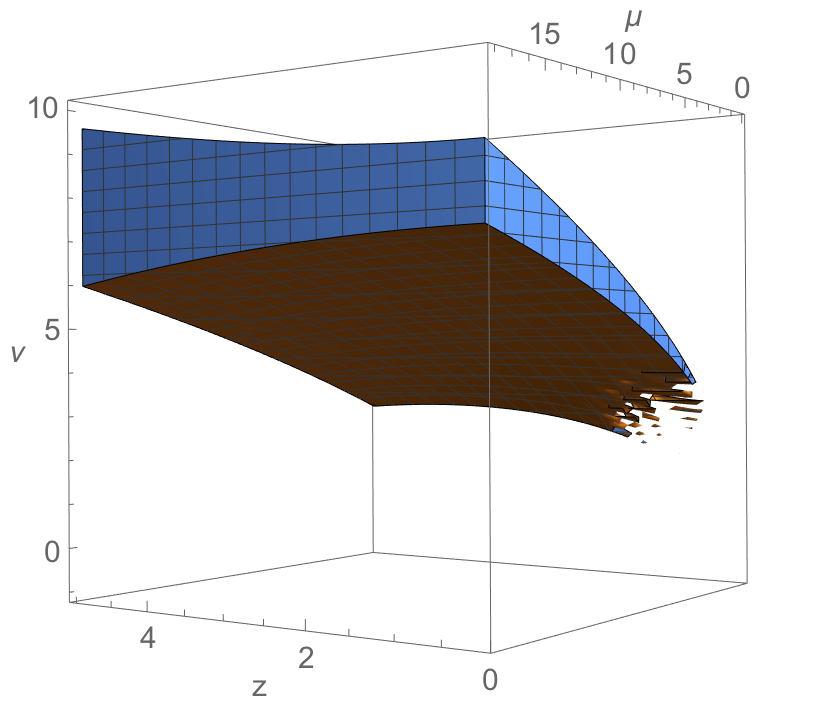}\includegraphics[width=.5\textwidth]{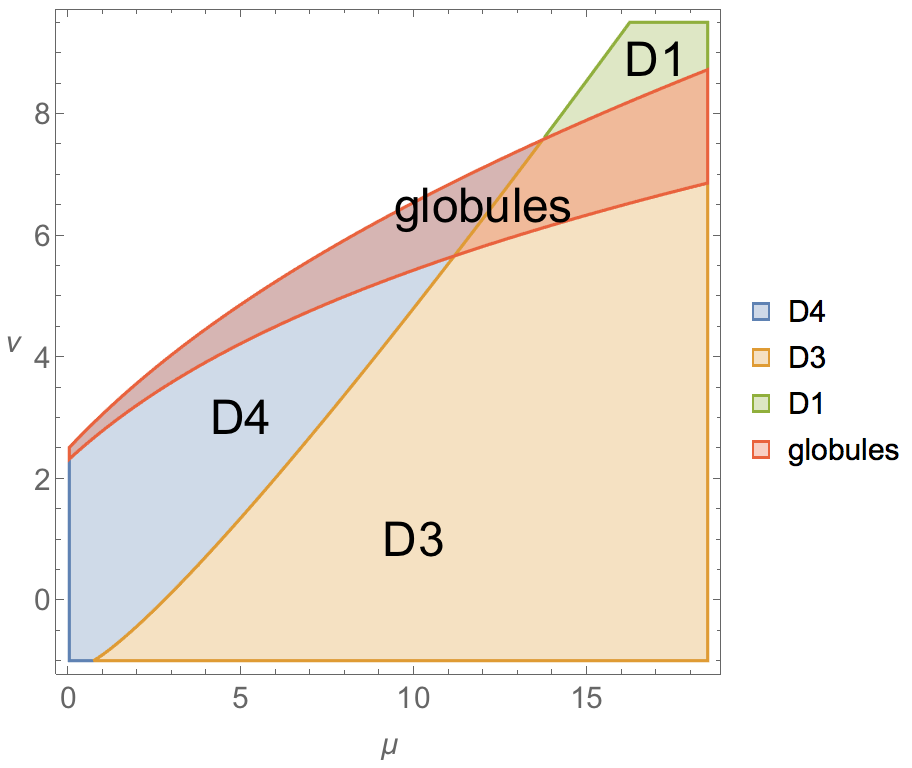}
\caption{\label{Fig26} On the left, the 3d plot of the region (blue) where black globules are present, and (on the right) a  section of the plot where the value of $z$ is fixed to $z=1.5$. One can see that increasing the charge $z$ the volume of the region where globules are present grows as well.}}
 \end{center}
\end{figure}

The three-dimensional plot in Fig \ref{Fig26} displays the region in parameter space where the globules are present, namely where the function $G(x)/x^2$ is not  monotonic. A two-dimensional section of this plot is also provided, for fixed $z$. One can see that increasing the charge increases the volume of the region where globules are present.

In region D3, there are charged versions of the droplet solution as studied in \cite{Caldarelli:2011wa}.

We end this section by pointing out that the charged analytic solutions discussed here exhibit a  puzzling behavior similar to the uncharged one described in sections \ref{RegD4} and 4.4, when approaching the  D3/D4 and D1/D2 boundaries. Indeed  a double root of $F(y)$ arises there, hence the boundary volume blows up due to the divergence of the integrand \eqref{regarea}. The horizon once again undergoes no apparent change. For the deformed spherical black holes the situation is analogous, and since also there the function $F(y)$ develops a double root at the border between D1 and D2, the area diverges as well. In all this, the area of the horizon  changes in a continuous way, without blowing up. The conclusion, as before, is that there must be new families of charged black holes with the same boundary geometry that will dominate the canonical ensemble.

\end{document}